\documentclass[12pt,nofootinbib,tightenlines,preprintnumbers,superscriptaddress,floatfix]{revtex4-1}
\usepackage[T1]{fontenc}
\usepackage[utf8]{inputenc}
\usepackage{lmodern}
\usepackage[english]{babel}
\usepackage{graphicx}
\usepackage{amsmath,mathtools,amssymb}
\usepackage{hyperref}
\usepackage{tabularx}
\usepackage{hhline}
\usepackage{booktabs}
\usepackage[dvipsnames]{xcolor}
\usepackage{tikz}
\usepackage{bm}
\usepackage{pgf-pie}

\usetikzlibrary{decorations.pathreplacing,decorations.pathmorphing}

\usepackage{siunitx}
\usepackage{soul}

\newcommand{\md}{\mathrm d}
\newcommand{\beq}{\begin{equation}}
\newcommand{\eeq}{\end{equation}}
\newcommand{\bit}{\begin{itemize}}
\newcommand{\eit}{\end{itemize}}
\newcommand{\ben}{\begin{enumerate}}
\newcommand{\een}{\end{enumerate}}

\usepackage{physics}
\renewcommand\bold{\mathbf}

\DeclareFontFamily{U}{mathx}{\hyphenchar\font45}
\DeclareFontShape{U}{mathx}{m}{n}{<-> mathx10}{}
\DeclareSymbolFont{mathx}{U}{mathx}{m}{n}
\DeclareMathAccent{\widebar}{0}{mathx}{"73}

\begin{document}

\preprint{MITP-22-098}

\title{Aspects of chiral symmetry in QCD at \texorpdfstring{$T=\SI{128}{\MeV}$}{T=128 MeV}}

\author{Marco C{\`e}}
\affiliation{Albert Einstein Center for Fundamental Physics (AEC)
and Institut für Theoretische Physik,
Universität Bern, Sidlerstrasse 5, CH-3012 Bern, Switzerland}

\author{Tim Harris} 
\affiliation{School of Physics and Astronomy,\\
University of Edinburgh, EH9 3JZ, UK}

\author{Ardit Krasniqi} 
\affiliation{PRISMA$^+$ Cluster of Excellence \& Institut f\"ur Kernphysik,
Johannes Gutenberg-Universit\"at Mainz,
D-55099 Mainz, Germany}

\author{Harvey~B.~\!Meyer} 
\affiliation{PRISMA$^+$ Cluster of Excellence \& Institut f\"ur Kernphysik,
Johannes Gutenberg-Universit\"at Mainz,
D-55099 Mainz, Germany}
\affiliation{Helmholtz~Institut~Mainz,
Johannes Gutenberg-Universit\"at Mainz,
D-55099 Mainz, Germany}
\affiliation{GSI Helmholtzzentrum f\"ur Schwerionenforschung, Darmstadt, Germany}

\author{Csaba T\"or\"ok} 
\affiliation{PRISMA$^+$ Cluster of Excellence \& Institut f\"ur Kernphysik,
Johannes Gutenberg-Universit\"at Mainz,
D-55099 Mainz, Germany}

\begin{abstract}
  We investigate several aspects of chiral symmetry in QCD at a temperature of $T = 128\,\text{MeV}$.
  The study is based on a $24\times 96^3$ lattice-QCD ensemble
  with O($a$)-improved Wilson quarks and physical up, down and strange quark masses.
  The pion quasiparticle turns out to be significantly lighter than the
  zero-temperature pion mass, even though the corresponding static correlation length is shorter.
  We perform a quantitative comparison of our findings to predictions of chiral perturbation theory.
  Among several order parameters for chiral symmetry restoration, we compute
  the difference of the vector- and axial-vector time-dependent correlators and find it to be
  reduced by a factor $\sim2/3$ as compared to its vacuum counterpart.

\end{abstract}

\maketitle

\section{Introduction}
\label{sec:intro}
Quark matter under extreme conditions (high temperatures and densities) is interesting both from the experimental and theoretical point of view. In the early universe (on a time scale of $\mathcal{O}(\mu s)$), the strongly interacting constituents (quarks and gluons) were in a hot and dense phase called {Quark-Gluon Plasma} (QGP).
Heavy ion colliders enable similar conditions to be reached in the lab. As a result of expansion, the universe gradually cooled down, undergoing a phase transition to a hadronic phase in which we now find ourselves. In the limit of massless quarks, the QCD Lagrangian has a global symmetry in flavor space, $SU(N_f)_L \times SU(N_f)_R$, corresponding to two independent rotations of the left- and right-handed components of the Dirac fields. This symmetry is spontaneously broken to $SU(N_f)_V$ and should be restored in the deconfined phase ({chiral symmetry restoration}). A non-vanishing value of the scalar density $\langle\bar{\psi}\psi\rangle(T)$ characterizes the low-temperature phase ($0\leq T\leq T_c$). On the contrary, $\langle\bar{\psi}\psi\rangle(T) = 0$ for $T>T_c$, indicating that chiral symmetry is restored. Thus, the quark  condensate $\langle\bar{\psi}\psi\rangle $ is a true order parameter for chiral symmetry breaking. A direct consequence of a restored chiral symmetry would be a coincidence of vector and axial-vector spectral function (see Sec.\,\ref{sec:Ioffe}).

Starting in the QCD vacuum, increasing the temperature initially leads to a dilute gas of pions. As the temperature is further increased, other hadron species also begin to contribute. At the same time, one expects the excitations of the medium to be quasiparticles with somewhat modified properties as compared to the standard hadrons, which are excitations of the vacuum.
A natural starting point in the investigation of the medium's quasiparticles is to examine the properties of the pion in the thermal environment \cite{Shuryak:1990ie,Goity:1989gs}. The pion mass and the pion decay constant have been studied to one loop in a thermal chiral perturbation theory (ChPT) approach \cite{Gasser:1987ah}. Additionally, the energy density, the pressure and the quark condensate have been investigated up to $\mathcal{O}(p^8)$ in a chiral expansion below the phase transition~\cite{Gerber:1988tt}. In Ref.\,\cite{Schenk:1993ru} the shift in the pion pole was calculated as a function of temperature up to second order in the density. Toublan~\cite{Toublan:1997rr} calculated also the pion decay constant within thermal ChPT to two loops and additionally examined the validity of the Gell-Mann--Oakes--Renner (GOR) relation at finite temperature. Unfortunately, it is unclear how far up in the temperature this expansion is applicable, since the partition function is certainly no longer dominated by the pions for $T \gtrsim 100$\,MeV~\cite{Shuryak:1990ie,Gerber:1988tt}. However, the Goldstone-boson nature of pions guarantees the existence of a divergent static correlation length  for vanishing quark masses~\cite{Pisarski:1996mt}.

In standard thermal ChPT, the quark mass as well as the temperature are treated as small parameters, resulting in an expansion around $m_q=0$ and $T=0$. In Refs.\,\cite{Son:2001ff,Son:2002ci}, however, Son and Stephanov investigated perturbations only around $m_q=0$, keeping the temperature $T$ fixed to any value in the chirally broken phase. Although an explicit relation of parameters like the quark condensate and pion decay constant to their $T=0$ counterparts is no longer possible in this framework, the validity of their results is extended to a regime where neither ChPT nor perturbative QCD is usable.
Since lattice simulations rely on the imaginary-time formalism and due to the lack of Lorentz invariance at finite temperature, extracting real-time observables such as ‘pole masses’ out of lattice quantities (e.g. Euclidean correlators) is a non-trivial task. Nevertheless, within the chiral effective theory approach of Son and Stephanov, the real part of the dispersion relation of soft pions can be obtained in terms of static Euclidean correlators. For the $N_{\mathrm{f}}=2$ case this has been done in~\cite{Brandt:2014qqa,Brandt:2015sxa}.

The paper is structured as follows: In Sec.\,\ref{sec:prelim} we start with the introduction of some basic definitions which have a key role in the description of the pion quasiparticle. We continue with the implementation of the lattice correlators, followed by a brief description of the numerical setup.  Our results, divided into subsections, are presented in Sec.\,\ref{sec:results}. First we extract the mass and decay constant of the screening pion (\ref{sec:PCAC}-\ref{sec:extraction_screening_quant}).
Next, we determine the pion velocity $u$ and examine its dependence on a finite pion thermal width $\Gamma(T)$ (\ref{sec:velocity}-\ref{sec:thermal_width}). Subsequently, we reconstruct a smeared and rescaled version of the axial spectral function and compare our results with the literature (\ref{sec:Backus}-\ref{sec:comparison_lit}). Thereafter, we compare our lattice estimate for the quark number susceptibility with the prediction from the hadron resonance gas model (HRG) in Sec.\,\ref{sec:suscep}. 
We also investigate in that section the effect of a modified dispersion
relation for the pion in the HRG.
Finally, we look at several order parameters for chiral symmetry restoration in Sec.\,\ref{sec:order_param} and give our conclusions in Sec.\,\ref{sec:conclusion}.

\section{Preliminaries}
\label{sec:prelim}
In this section we introduce the notation and some basic definitions as well as the key quantities for the pion quasiparticle that we will use throughout the paper. Furthermore, the lattice implementation of the correlators and the numerical setup are described briefly.

\subsection{Definition of operators and correlation functions}
\label{sec:basic definitions}

The notation and conventions used in this work are adapted from Ref.\,\cite{Brandt:2014qqa}. Our framework is the light-quark sector of Euclideanized QCD on the space $S^1 \times \mathbb{R}^3$, $S^1$ denoting the Matsubara cycle of length $\beta \equiv 1/T$. We define the pseudoscalar density, the vector current and the axial-vector current as
\begin{align}
    \label{eq:ps. density, vector curr, ax. curr}
    P^a(x) = \bar{\psi}(x)\gamma_5\frac{\tau^a}{2}\psi(x)\,,\ \ \ V^a_{\mu}(x) = \bar{\psi}(x)\gamma_{\mu}\frac{\tau^a}{2}\psi(x)\,,\ \ A^a_{\mu}(x) = \bar{\psi}(x)\gamma_{\mu}\gamma_5\frac{\tau^a}{2}\psi(x)\,,
\end{align}
where $a \in \{1,2,3\}$ is an adjoint $SU(2)_{\text{isospin}}$ index, $\tau^a$ is a Pauli matrix and $\psi(x)$ is a Dirac field flavor doublet. The partially conserved axial current (PCAC) relation is an operator identity that holds in Euclidean space when inserted in expectation values. It relates the divergence of the axial vector current $A^a_{\mu}(x)$ to the pseudoscalar density $P^a(x)$,
\begin{align}
    \label{eq:PCAC relation}
    \partial_{\mu}A_{\mu}^a(x) = 2m_{\text{PCAC}}\,P^a(x)\,.
\end{align}
In the path integral formulation, this relation results from performing a chiral rotation $\delta_A^a\psi(x)=\frac{\tau^a}{2}\gamma_5\psi(x)$ of the fields (see Ref.\,\cite{Luscher:1998pe}). Applying the pseudoscalar density operator on both sides and taking the expectation value one can solve for the bare PCAC quark mass,
\begin{align}
    \label{eq:PCAC bare mass}
    m_{\text{PCAC}} = \frac{1}{2}\frac{\partial_{\mu}\langle A_{\mu}^a(x)P^a(0)\rangle}{\langle P^b(x)P^b(0)\rangle}\,.
\end{align}
 Since the PCAC relation is an operator identity, we are free to choose the direction in which we define the quark mass. 
 In our thermal system, the spatial direction is four times larger than the temporal one. As a consequence, measuring along the spatial direction results in a longer plateau and thus, smaller errors.
 Therefore, we will extract the PCAC quark mass from the relation
 \begin{align}
   \label{eq:mPCAClat}
    m_{\text{PCAC}}(x_3) = \frac{1}{2}\frac{\int\mathrm{d}x_0\mathrm{d}^2x_{\perp}\,\langle\widetilde\partial_3 A_3^{a}(x)P^a(0)\rangle}{\int\mathrm{d}x_0\mathrm{d}^2x_{\perp}\,\langle P^b(x)P^b(0)\rangle}\, ,\ \ \ \ \ \ x_{\perp} = (x_1,x_2)\, .
\end{align}

We introduce the static screening axial correlator,  given by
\begin{align}
    \label{eq:asymp}
    \delta^{ab} G_A^s(x_3,T) = \int \, \md x_0 \md ^2x_{\perp} \langle A_3^{a}(x)A_3^{b}(0) \rangle \overset{\abs{x_3}\rightarrow \infty}{=} \delta^{ab} \frac{f_{\pi}^2m_{\pi}}{2} e^{-m_{\pi} \abs{x_3}}\, ,
\end{align}
where we have specified the asymptotic form of the correlator,
which defines the screening pion mass $m_\pi$ and decay constant $f_\pi$.
Analogously, we define the following static screening static screening correlators:
\begin{align}
    \label{eq:def_static_PP_and_AP_corr}
    \delta^{ab} G_P^s(x_3,T) &= \int \, \md x_0 \md ^2x_{\perp} \langle P^a(x)P^b(0) \rangle \\
    \delta^{ab} G_{AP}^s(x_3,T) &= \int \, \md x_0 \md ^2x_{\perp} \langle A_3^{a}(x)P^b(0) \rangle\,.
\end{align}
The PCAC relation [see Eq.\,(\ref{eq:PCAC relation})] implies the relation
 \beq\label{eq:GPsGAs}
G_P^s(x_3,T) = - \frac{1}{{4m_{\text{PCAC}}^2}} \frac{\partial^2}{\partial x_3^2}  G_A^s(x_3,T).
 \eeq

 In order to probe the dynamical properties of the thermal system,
we define time-dependent correlators, projected to a definite spatial momentum,
\begin{align}
  \delta^{ab} G_{A_0}(x_0,T) &= \int \, \md^3x\, \langle A_0^{a}(x)A_0^{b}(0) \rangle \\
    \delta^{ab} G_{P}(x_0,T) &= \int \, \md^3x\, \langle P^a(x)P^b(0) \rangle \\
    \delta^{ab} G_{PA_0}(x_0,T) &= \int \, \md^3x\, \langle P^a(x)A_0^{b}(0) \rangle \\
    \label{eq:def_time_dep_corr}
    \delta^{ab} G_{A}(x_0,{\bf p},T) &= -\frac{1}{3} \sum_{i=1}^3 \int \, \md^3x\, e^{-i{\bf p}\cdot {\bf x}}\,\langle A_i^{a}(x)A_i^{b}(0) \rangle \\
    \delta^{ab} G_{V}(x_0,{\bf p},T ) &= -\frac{1}{3} \sum_{i=1}^3 \int \, \md^3x\, e^{-i{\bf p}\cdot {\bf x}}\,\langle V_i^{a}(x)V_i^{b}(0) \rangle\,.
\end{align}

The time-dependent axial correlator $G_{A}(x_0,T)$ [see Eq.\,(\ref{eq:def_time_dep_corr})] can be obtained from the spectral function $\rho_A(\omega, T, \bold{p})$ (see e.g.\ the review~\cite{Meyer:2011gj}):
\begin{equation}
  \label{eq:axial_spec_func}
  G_A(x_0,T,\bold{p}) = \int_0^{\infty}\,\mathrm{d}\omega\, \rho_A(\omega, {\bf p},T)\,\frac{\text{cosh}(\omega(\beta/2-x_0))}{\text{sinh}(\omega\beta/2)}\,.
\end{equation}
In Sec.\,\ref{sec:Backus} we will analyze the axial spectral function using the Backus-Gilbert method.
Relations analogous to Eq.\,(\ref{eq:axial_spec_func}) hold for the correlators $G_{A_0}(x_0,T)$, $G_{P}(x_0,T)$ and $G_{V}(x_0,{\bf p},T)$.

\subsection{Pion properties at finite temperature}
It has been established within several frameworks~\cite{Son:2001ff,Son:2002ci} that at temperatures well below the chiral phase transition a pion quasiparticle persists, with the real part of the dispersion relation of sufficiently soft pions given by
\begin{align}
    \label{eq:dispersion}
    \omega_{\bold{p}} = u(T) \sqrt{m_{\pi}^2 + \bold{p}^2} \, , \ \ \ \text{for any} \  T \lesssim T_c\, .
\end{align}
In the chiral limit it can be interpreted as the group velocity of a massless pion excitation.
While the quasiparticle mass $\omega_{\bold{0}}$ is the real-part of a pole of the retarded correlator $G^R_P(\omega, {\bf p}=0, T)$
of the pseudoscalar density in the frequency variable, the screening mass $m_\pi$ is a pole of
$G^R_P(\omega = 0, \bold{p},T)$ in the spatial momentum $\abs{\bold{p}}$ and represents an inverse spatial correlation length.
A simple interpretation of the dispersion relation (\ref{eq:dispersion}) was given in Ref.\,\cite{Brandt:2015sxa} in terms of the poles
of the screening and the time-dependent correlators. Son and Stephanov~\cite{Son:2001ff,Son:2002ci}
showed that the pion velocity $u$ in the chiral limit is the ratio of two static  quantities,
\begin{align}
    \label{eq:u_Son}
    u^2 = \frac{f_{\pi}^2}{\int_0^{\beta} \md x_0\, G_{A_0}(x_0,T)}\,.
\end{align}
As noted in Ref.\,\cite{Brandt:2014qqa}, the axial susceptibility appearing in the denominator of Eq.\,(\ref{eq:u_Son}) contains an ultraviolet divergence at any non-vanishing quark mass and is therefore not practical for lattice calculations. As an alternative, in Refs.\,\cite{Brandt:2014qqa,Brandt:2015sxa} the parameter $u$ was determined using lattice correlation functions at vanishing spatial momentum via the two estimators,
\begin{align}
    \label{eq:u_m}
    u_m &= \left[\left.-\frac{4m_q^2}{m_{\pi}^2}\,\frac{G_P(x_0,T)}{G_{A_0}(x_0,T}\right\vert_{x_0=\beta/2}\right]^{1/2}\, , \\
    \label{eq:u_f}
    u_f &= \frac{f_{\pi}^2 m_{\pi}}{2G_{A_0}(\beta/2,T)\,\text{sinh}(u_fm_{\pi}\beta/2)}\, ,
\end{align}
which we will adopt. In doing so, for the estimator $u_m$, the parametric dominance of the pion in the time-dependent Euclidean axial as well as the pseudoscalar density correlator at small quark masses is exploited. The estimator $u_f$ exploits only the parametric dominance of the axial correlator; on the other hand, it relies on the residue determined from the static screening correlator.
The pion contribution to the spectral function $\rho_{A_0}$ is expected to take the form of a sharp peak, 
\begin{equation}
\label{app:eq:spec}
    \rho_{A_0}(\omega, T) = \text{sgn}(\omega)\text{Res}(\omega_{\bold{0}})\delta(\omega^2-\omega_{\bold{0}}^2)+\dots\,,
\end{equation}
where in Ref.\,\cite{Son:2002ci} (see also Ref.\,\cite{Brandt:2015sxa}) the residue was predicted to have the form 
\begin{equation}
    \label{eq:def_res}
 \text{Res}(\omega_{\bold{0}}) \equiv (f_{\pi}^t)^2\omega_0^2 = f_{\pi}^2m_{\pi}^2\,,
\end{equation}
such that we can access the quasiparticle decay constant via $f_{\pi}^t = f_{\pi}/u_m$.

\subsection{Lattice implementation of the correlators}
\label{sec:lat_impl}
In this work we use exclusively the local discretizations of the operators introduced in the previous subsection.
Therefore, the expression of the bare operators in the lattice theory is the same as in Eq.\,(\ref{eq:ps. density, vector curr, ax. curr}).
These bare operators are first O($a$)-improved and then renormalized.
While the bare pseudoscalar density is by itself O($a$)-improved, the improvement of the vector and axial-vector currents takes the form
\begin{align}
\label{eq: imp_axial_corr}
A_{\mu}^{\text{imp},b}(x) &= A_{\mu}^b(x) + ac_A(g_0^2)\widetilde\partial_{\mu} P^b(x)\, ,
\\
    \label{eq:imp_vec}
    V^{\text{imp},b}_{\mu}(x) &= V_{\mu}^b(x) + ac_V(g_0^2)\widetilde\partial_{\nu}\,T_{\mu\nu}(x)\, ,
\end{align}
where $T_{\mu\nu}^a(x)\equiv - \frac{1}{2} \bar\psi [\gamma_\mu,\gamma_\nu]\frac{\tau^a}{2}\psi$ is the tensor current.
For the derivative $\widetilde\partial_{\mu}$, we use the symmetrized version of the lattice discretized derivative, namely 
\begin{align}
   \label{eq:symm_deriv}
    \widetilde\partial_{\mu} f(x) = \frac{1}{2a}\left(f(x+a\hat{\mu}) - f(x-a\hat{\mu})\right)\, .
\end{align}
The non-perturbatively calculated coefficient $c_A$ was taken from Ref.\,\cite{Bulava:2015bxa},
and the coefficient $c_V$ from Ref.\,\cite{Gerardin:2018kpy}. 

The finite renormalization of the vector and the axial-vector currents
is performed with the non-perturbatively determined renormalization
factors $Z_V(g_0^2)$ and $Z_A(g_0^2)$, supplemented by a quark-mass
dependent factor in order to fully realize O($a$) improvement; details
are provided in Appendix~\ref{app:ren}.

The pseudoscalar density $P^a(x)$ acquires a scale (and scheme) dependence via the process of renormalization.
The renormalization factor is notated $Z_P(g_0^2,a\mu)$.
Here, we renormalize $P^a(x)$ in the (non-perturbative)
gradient-flow (GF) scheme at the renormalization scale $\mu$ where the corresponding coupling $\bar g^2_{\rm GF}=9.25$;
this corresponds to a low scale of  $\mu\approx230\;$MeV \cite{Campos:2018ahf}.
While none of our physics applications relies on the choice of a specific scheme, we note that in the latter publication,
the scale dependence of the renormalization factor has been computed up to perturbative scales $\mu$;
thereby the connection to the renormalization-group invariant (RGI) operator is known.

The renormalization of the PCAC mass is defined to preserve the axial Ward identity (\ref{eq:PCAC relation}). Thus, all renormalization-scale dependent quantities in this paper are quoted in the aforementioned gradient-flow scheme.
In particular the PCAC mass is renormalized by multiplying it with $Z_A/Z_P$, and the combination $m_\pi^2 f_\pi^2 / m_{\rm PCAC}$ considered
in Sec.\,\ref{sec:GOR} via the factor $ Z_A Z_P$. The numerical values of the renormalization factors are collected in Appendix~\ref{app:ren}.

\subsection{Numerical setup}

Our calculations are performed on an $N_{\mathrm{f}}=2+1$ ensemble with tree-level $\mathcal{O}(a^2)$-improved Lüscher-Weisz gauge action and non-perturbatively $\mathcal{O}(a)$-improved Wilson fermions\,\cite{Bulava:2013cta}.
The action corresponds to the choice of the Coordinated Lattice Simulations (CLS) initiative\,\cite{Bruno:2014jqa} and the bare parameters match those of the CLS zero-temperature ensemble E250\,\cite{Mohler:2017wnb}.
The latter are listed in Table\,\ref{tab:E250params}, together with the lattice spacing as determined in Ref.\,\cite{Bruno:2016plf}. 
The time direction admits thermal boundary conditions with $N_{\tau}=24$, which is the only difference relative to the zero-temperature ensemble, resulting in a temperature 
\beq
T = \frac{1}{\beta} = \frac{1}{24a} = 127.9(1.5)\,{\rm MeV}.
\eeq
Note that, assuming a critical temperature $T_c  \approx 150$\,MeV in $(2+1)$-flavor QCD \cite{Aoki:2006br, Aoki:2009sc,Borsanyi:2013bia}, our temperature corresponds to $T/T_c \approx 0.85$.
For reference, we also quote the zero-temperature pseudoscalar masses and pion decay constant, determined in Ref.\,\cite{Ce:2022kxy},
\begin{align}
\label{eq:mpimKT0}
T=0:\qquad  m_\pi^0 &= 128.1(1.3)(1.5) \,{\rm MeV}, \qquad m_K^0 = 488.98(0.3)(5.8)\, {\rm MeV},\\
f_{\pi}^0 &= 87.4(0.4)(1.0)\,{\rm MeV}
\end{align}
where the first error is from the corresponding quantity in lattice units, and the second is from
the lattice spacing determination of Ref.\,\cite{Bruno:2016plf}.

\begin{table}[tb]
\caption{\label{tab:E250params} Parameters and lattice spacing of the ensemble analyzed in this paper.
    The lattice spacing determination is from Ref.\,\cite{Bruno:2016plf}.
  }
  \begin{tabular}{c@{~~~}c@{~~~}c@{~~~}c@{~~~}c@{~~~}c}
    \hline
    \hline
    $\beta/a$ & $L/a$  & $6/g_0^2$ & $\kappa_l$ & $\kappa_s$ & $a\,[{\rm fm}]$ \\
    \hline
    24  & 96  & 3.55  & 0.137232867 & 0.136536633  &  0.06426(76) \\
    \hline
    \hline
  \end{tabular}
  \end{table}

The ensemble has been generated using version 2.0 of the openQCD package\,\cite{Luscher:2012av}, applying a small twisted mass to the light quark doublet for algorithmic stability.
The correct QCD expectation values are obtained including the reweighting factors\footnote{We have not found any negative reweighting factors.} for the twisted mass and for the RHMC algorithm approximation used to simulate the strange quark.
Measurements are performed on a single chain of 1200 configurations, each separated by four MDUs.

\section{Results on the pseudoscalar sector}
\label{sec:results}

In this section, we present our lattice results on observables in the pseudoscalar sector, i.e.\ those related to pion properties.
As an important reference quantity, we begin with the determination of the average $(u,d)$ PCAC mass.

\subsection{The PCAC mass}
\label{sec:PCAC}

\begin{figure}[tp]
	\includegraphics[scale=0.80]{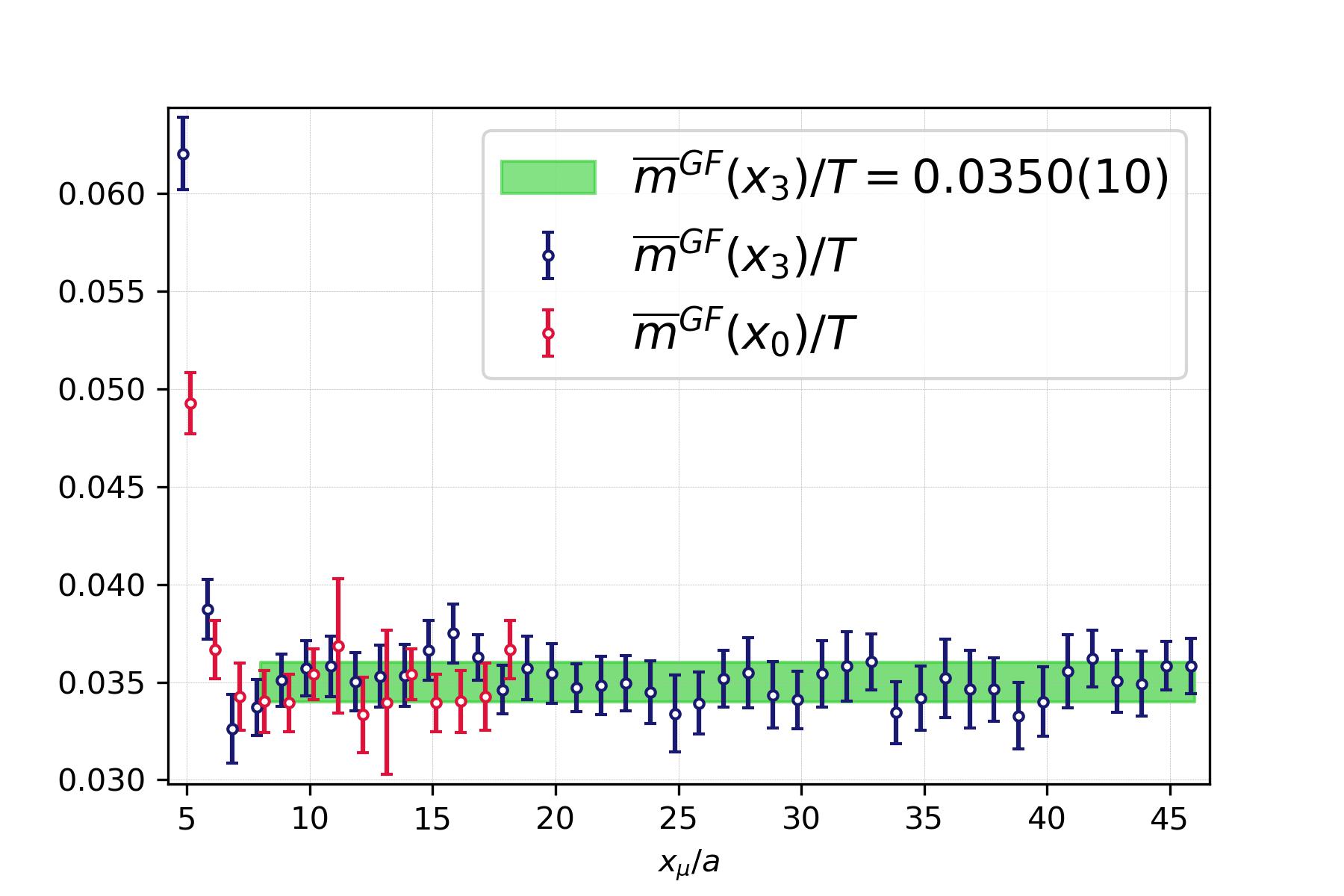}
	\caption{Renormalized PCAC mass in the E250 ensemble along the $x_3$-direction. The final result --- obtained from a fit along the $x_3$-direction --- is also shown with a $1 - \sigma$ band. We have used the improved axial current together with the symmetrized derivative (see Eqs.\,(\ref{eq: imp_axial_corr}-\ref{eq:symm_deriv})).
	\label{fig:PCAC}}
\end{figure}

The extraction of the PCAC mass as defined in Eq.\,(\ref{eq:mPCAClat})
is carried out by performing a fit to a constant in the range where a plateau is observed; see Fig.\,\ref{fig:PCAC}.
Due to the longer plateau, the fit is performed along the $x_3$- direction. We obtain
\begin{equation}
    \frac{m_{\text{PCAC}}}{T} = 0.035(1)
\end{equation}
The PCAC-mass obtained from the $x_0$-direction is compatible with the one obtained from the $x_3$-direction, pointing to cutoff effects at this value of the lattice spacing being small.

\subsection{Static correlators: the pion screening mass and decay constant}
\label{sec:extraction_screening_quant}

\begin{figure}[tp]
	\includegraphics[scale=0.70]{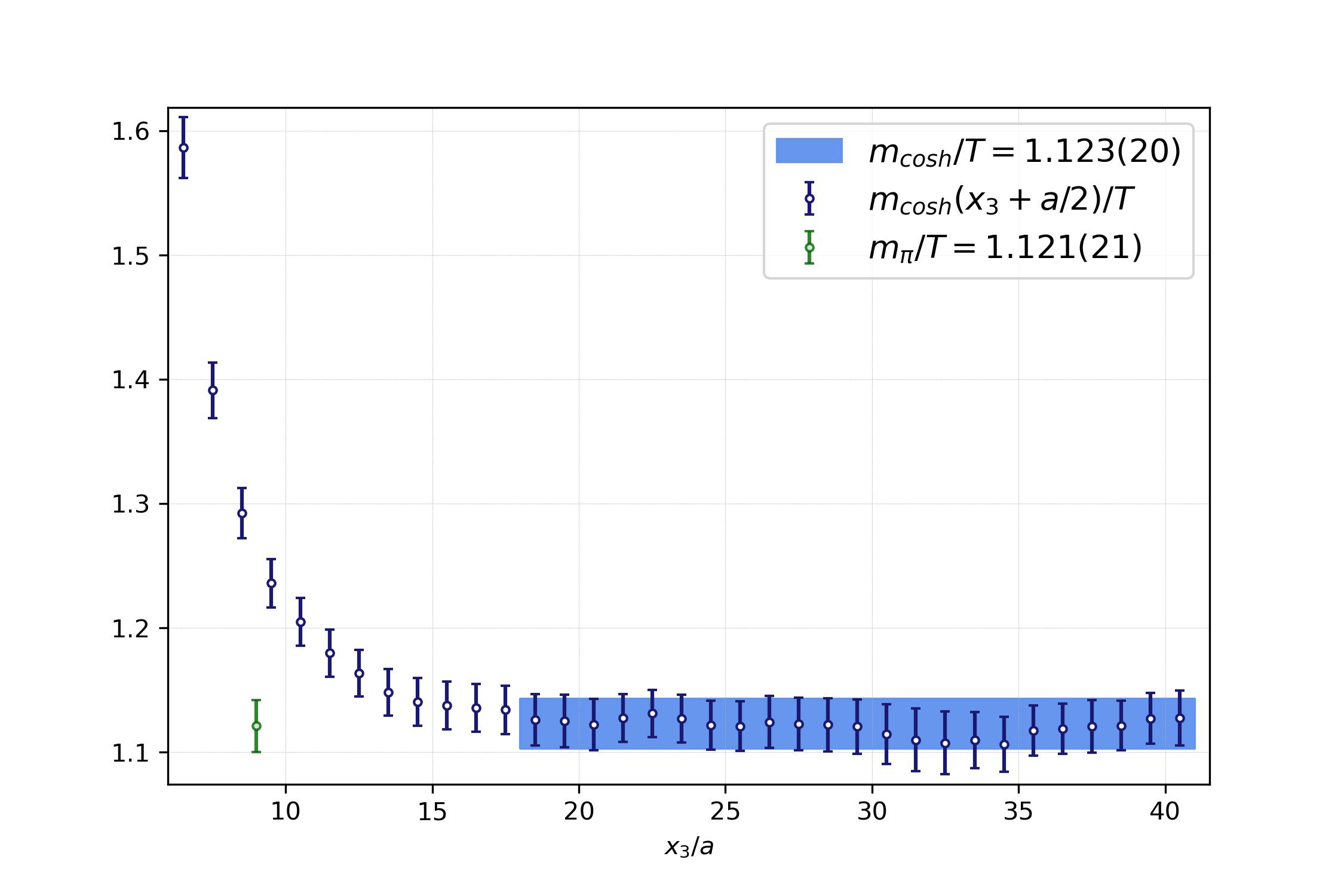}
	\caption{Effective mass plot for the cosh mass $m_{\text{cosh}}(x_3)$ as a function of the $x_3$-coordinate, obtained from the pseudoscalar screening correlation function at zero spatial momentum $G_P^s(x_3,T)$. It is assumed that the effective mass plateau starts at $x_3/a = 18$. The result of the fit to the effective mass values is represented by a $1-\sigma$ band. For comparison the value for the screening pion mass $m_{\pi}$, obtained from the fit of the pseudoscalar correlator is also included. 
	\label{fig:cosh}}
\end{figure}

In this subsection, we describe how the screening pion mass $m_{\pi}$ and the screening decay constant $f_{\pi}$ can be calculated. In order to accomplish this, we make use of the asymptotic behavior of the axial-current screening correlator, Eq.\,(\ref{eq:asymp}).

Making use of the PCAC-based relation (\ref{eq:GPsGAs}) and of the symmetry of the correlators 
around $x_3 = L/2$, a one-state fit ansatz for the corresponding correlation functions can be formulated in the form 
\begin{align}
  \label{eq:axial_fit}
  G_A^s(x_3,T) &= \frac{A_1^2 m_1}{2}\, \text{cosh}[(m_1(x_3-L/2))]\, , \\
  G_P^s(x_3,T) &= -\frac{A_1^2 m_1^3}{8m_{\text{PCAC}}^2}\, \text{cosh}[(m_1(x_3-L/2))]\, , \\
  G_{AP}^s(x_3,T) &= -\frac{A_1^2 m_1^2}{4m_{\text{PCAC}}}\, \text{sinh}[(m_1(x_3-L/2))]\, .
\end{align}
The pion screening mass $m_\pi$ and $f_\pi$ are obtained from the fit parameters $m_1$ and $A_1$ (derived in Appendix \ref{app:extracting}) via
\beq
m_\pi = m_1, \qquad f_\pi = A_1\,\sqrt{\sinh(m_1 L/2)}\;.
\eeq

\begin{figure}[tp]
	\includegraphics[scale=0.58]{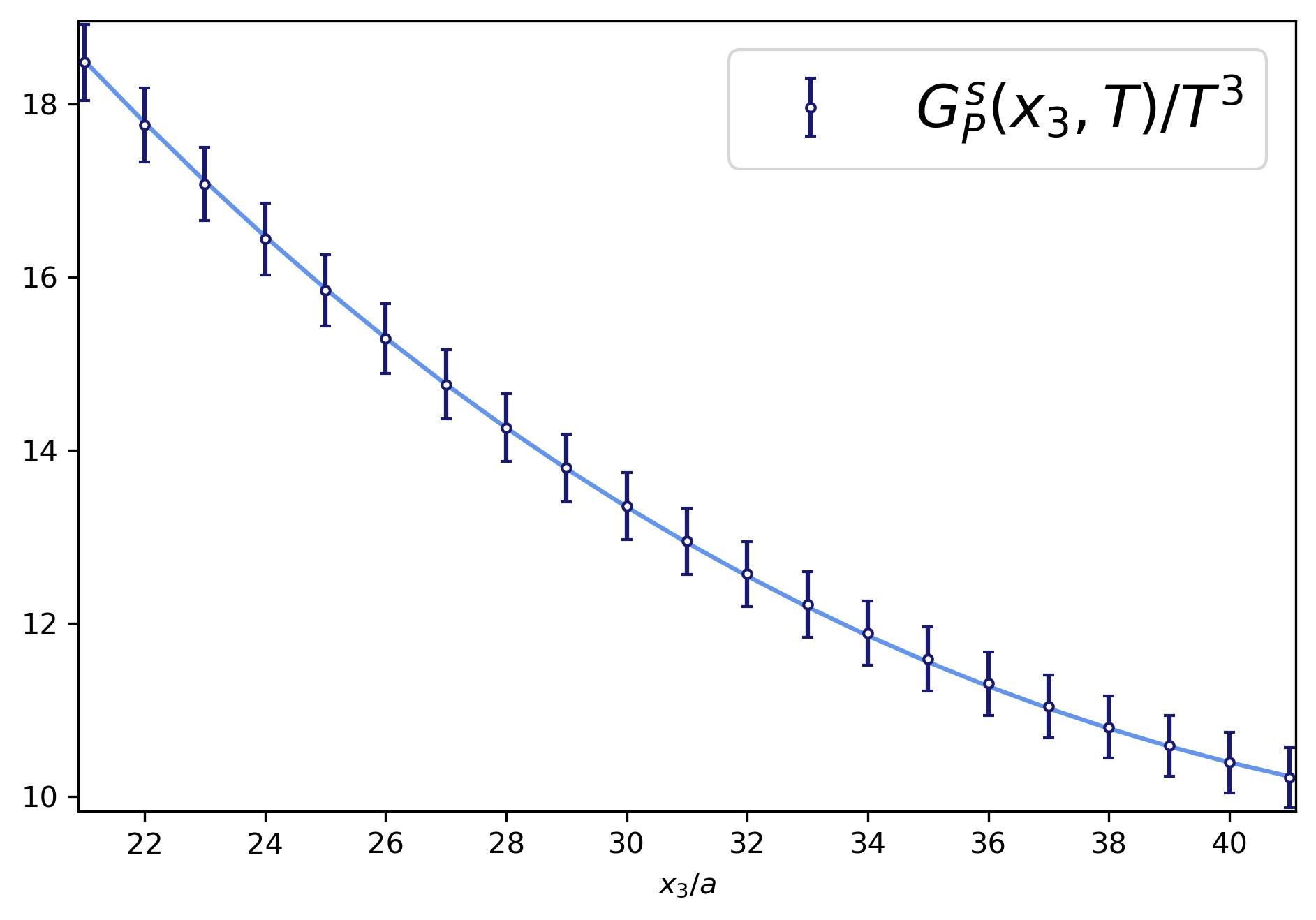}  \hspace{0.5mm}
	\includegraphics[scale=0.58]{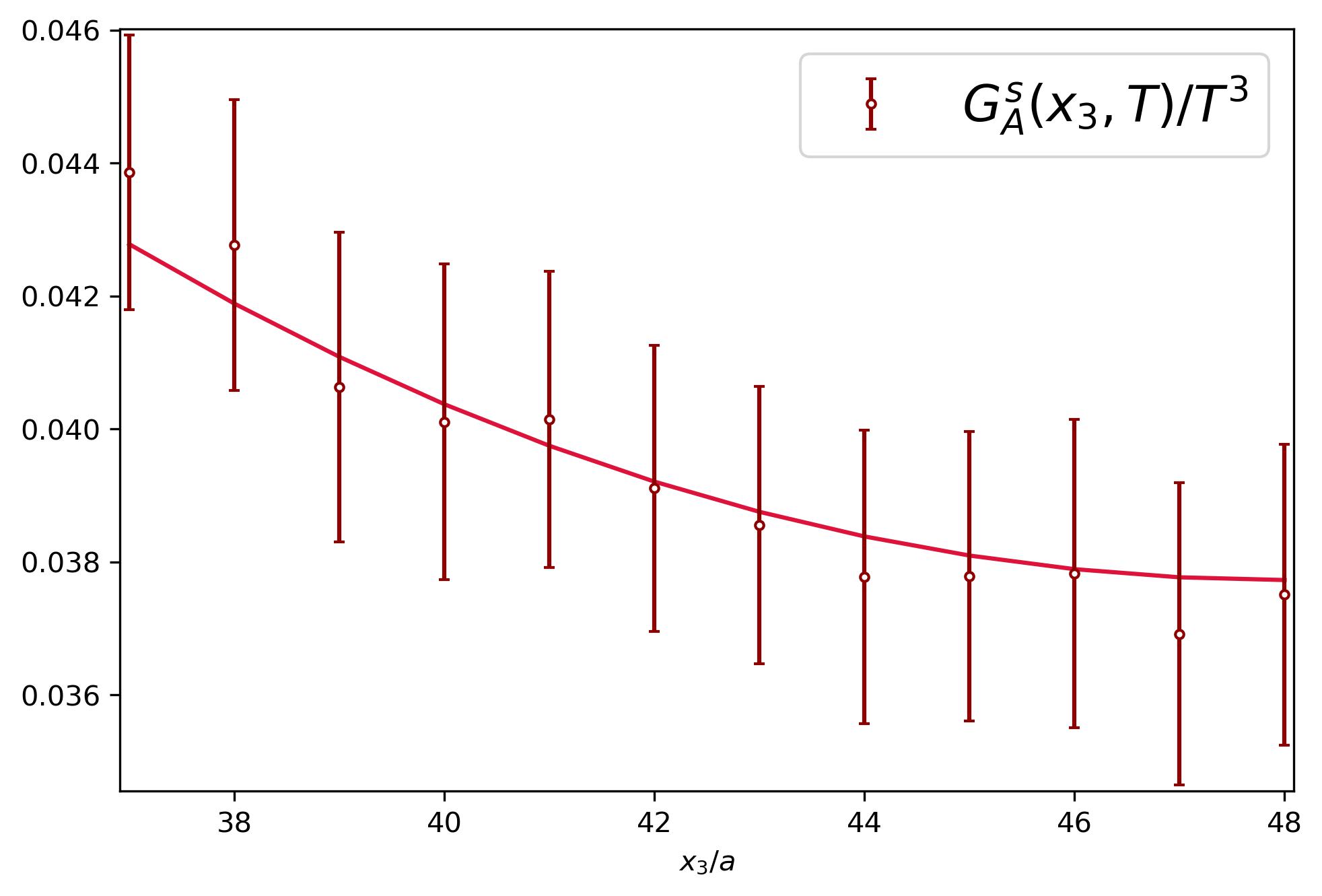}
    \includegraphics[scale=.58]{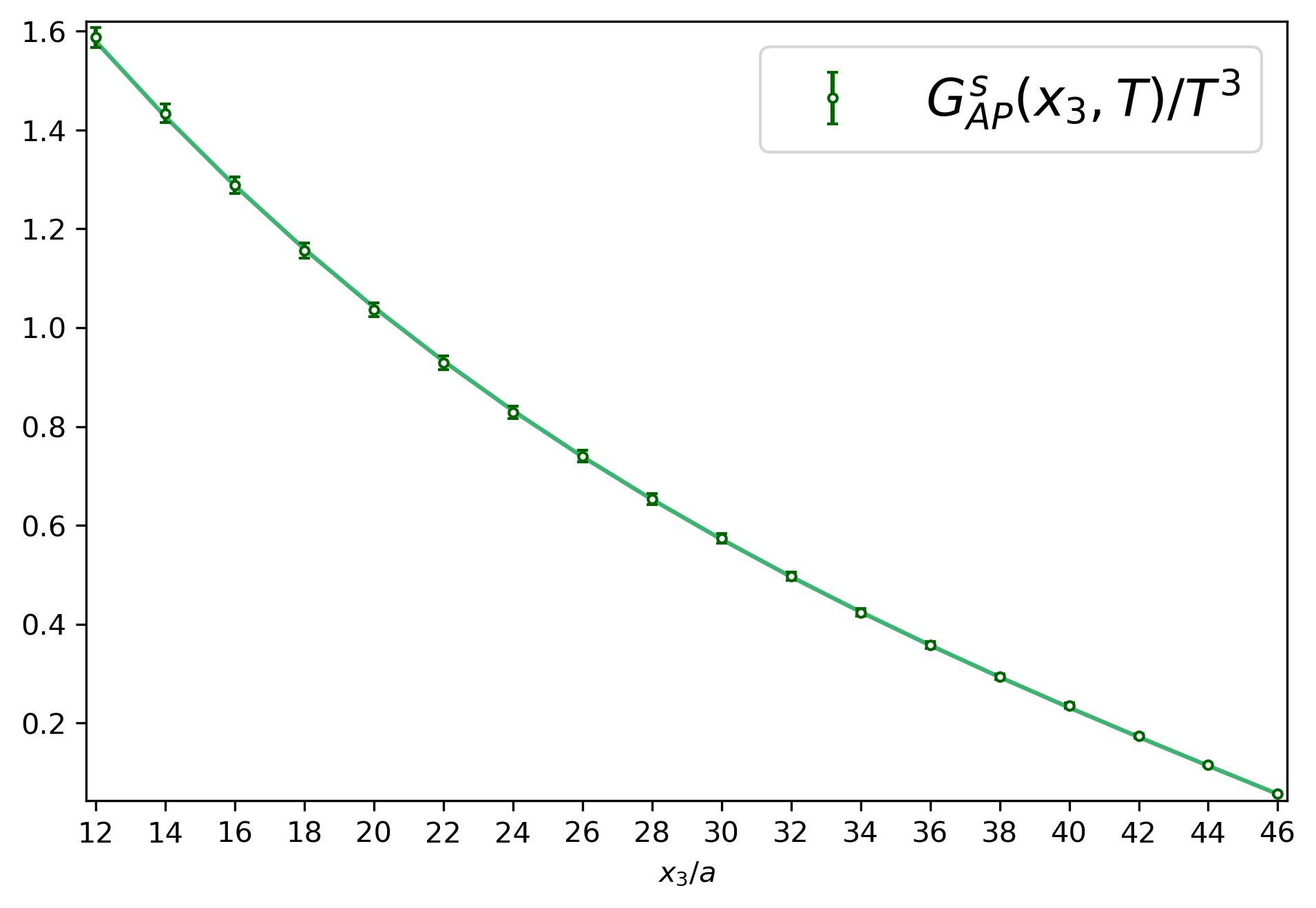}
\caption{{\bfseries Top panel}: Renormalized screening correlation function $G_P^s(x_3,T)/T^3$ and the result of the fit. The chosen fit interval is $x_3/a \in [21,41]$.
{\bfseries Middle panel}: Renormalized screening correlation function $G_A^s(x_3,T)/T^3$ and the result of the fit with a prior from $G_P^s(x_3,T)$. The chosen fit interval in this case is $x_3/a \in [37,48]$. {\bfseries Bottom panel}: Renormalized screening correlation function $G_{AP}^s(x_3,T)/T^3$ and the result of the fit with a prior from $G_P^s(x_3,T)$. The chosen fit interval in this case is $x_3/a \in [12,46]$.}
\label{fig:corr}
\end{figure}

The `cosh mass’ with argument $(x_3+a/2)$ is defined as the positive root of the following equation,
\begin{align}
    \frac{G_P^s(x_3,T)}{G_P^s(x_3+a,T)} = \frac{\text{cosh}[m_{\text{cosh}}(x_3+a/2)\cdot(x_3-L/2)]}{\text{cosh}[m_{\text{cosh}}(x_3+a/2)\cdot(x_3+a-L/2)]}\,.
\end{align}
It is visualized in Fig.\,\ref{fig:cosh}. Note that there is a different equation and a different solution for $m_{\text{cosh}}$ for each value of $x_3$.
The fits are performed using the Levenberg-Marquardt’s method \cite{press2007numerical}
and the results for the pion screening mass are shown in Fig.\,\ref{fig:corr}.

Due to its better signal-to-noise ratio, the screening pion mass $m_{\pi}$ was first extracted using $G_P^s(x_3,T)$. Since neighbouring correlator points are highly correlated, we have fitted only every second point. Proceeding in this way, the dimension of the covariance matrix is reduced, enabling us to perform correlated fits over a longer physical range of distances.  In order to be sure that the ground state is isolated, we have repeated the fit to the correlation function for different fit windows, leaving out points that are furthest away from the correlator middle point $x_3 = L/2$. Our final result
\begin{equation}
    m_{\pi}/T = 1.121(21)\,,
\end{equation}
corresponding to $m_{\pi} = 144(3)$\,MeV. It is
reported in Table\,\ref{tab:results} and is stable under small variations of the fit interval and corresponds to a correlated $\chi^2/\text{d.o.f.} = 1.05$, where $\text{d.o.f.} = 9$. Furthermore, our final value for $m_{\pi}$ is in very good agreement with the averaged value  of the `cosh mass’ $m_{\text{cosh}} = 1.123(20)$; see Fig.\,\ref{fig:cosh}. The obtained fit parameters of $G_P^s(x_3,T)$ are then used as a prior to fit  $G_A^s(x_3,T)$ and $G_{AP}^s(x_3,T)$ .

By repeating the procedure for different fit windows in an analogous manner, we note that the mean value is stable under small variations of the fit window and we select the final value for $A_1$ by choosing a fit which has a correlated $\chi^2/\text{d.o.f.} = 1.14$ ($\text{d.o.f.} = 9$) for $G_A^s(x_3,T)$ and $\chi^2/\text{d.o.f.} = 0.98$ ($\text{d.o.f.} = 15$) for $G_{AP}^s(x_3,T)$. Employing Eq.\,(\ref{eq:decay_const}) the value $f_{\pi}/T = 0.558(14)$, respectively $f_{\pi}/T = 0.559(11)$  can be extracted for the screening pion decay constant. The latter value is selected as the final one and reported in Table\,\ref{tab:results}. 
The screening pion decay constant $f_{\pi} = 72(1)$\,MeV  is significantly lower than the pion decay constant $f_{\pi}^0 = 87.4(1.0)$\,MeV \cite{Ce:2022eix} on the corresponding zero temperature ensemble\footnote{Note that in this work a different convention for the pion decay constant is followed resulting in an additional factor $\sqrt{2}$.}
The procedure for calculating the statistical error of the screening quantities $m_{\pi}$ and $f_{\pi}$ is described in Appendix \ref{app:autocorr}.

\subsection{Properties of the pion quasiparticle}
\label{sec:velocity}

The results for the estimators $u_m$ and $u_f$ defined in Eqs.\,(\ref{eq:u_m},\ref{eq:u_f})
together with the estimators for the screening quantities are presented in Table\,\ref{tab:results}.
Good agreement is found for the two independent estimators $u_f$ and $u_m$ of the pion velocity. Both of them differ significantly from unity,
which clearly represents a breaking of Lorentz invariance due to thermal effects.
Additionally, we found that the zero-temperature pion mass given in Eq.\,(\ref{eq:mpimKT0}) `splits' into a lower pion quasiparticle mass, $\omega_{\bold{0}} = 113(3)$\,MeV, and a higher pion screening mass, $m_{\pi} = 144(3)$\,MeV.
The quasiparticle decay constant $f^{t}_{\pi} = f_{\pi}/u_m = 91(2)$\,MeV is much closer to the vacuum decay constant $f_{\pi}^0$.

\begin{table}[tb]
\caption{Summary of the results of the E250 thermal ensemble with $N_{\tau} = 24$. The pion quasiparticle mass $\omega_{\bold{0}}$ is calculated using $\omega_{\bold{0}} = u_m m_{\pi}$.} 
\begin{tabular}{lS[table-format=1.3(2)]}
\hline
\hline
$m_{\pi}/T$                         & 1.121(21) \\ 
$f_{\pi}/T$                         & 0.559(11)\\ \hline
$u_{f}$                             & 0.787(16) \\
$u_{m}$                             & 0.786(18) \\
$u_{f}/u_m$                         & 1.001(27) \\ \hline
$\omega_{\bold{0}}/T$               & 0.881(23) \\
$f_{\pi}^t/T$                       & 0.710(16) \\
$\text{Res}(\omega_{\bold{0}})/T^4$ & 0.392(21) \\
\hline
\hline
\end{tabular}
\label{tab:results}
\end{table}

\subsection{Dependence of the pion velocity \texorpdfstring{$u_f$}{uf} on a finite pion thermal width \texorpdfstring{$\Gamma(T)$}{Γ(T)}}
\label{sec:thermal_width}

The analysis of Son and Stephanov~\cite{Son:2001ff, Son:2002ci} concluded that at temperatures below the chiral phase transition, the  imaginary part of the pion pole is parametrically small compared to its real part.
In this subsection, we investigate the sensitivity of our results for the pion quasiparticle mass and velocity parameter $u$
to the assumption of a negligible thermal width of this quasiparticle.
In order to examine the consequences of a finite thermal pion width on the pion velocity $u$ we replace the $\delta$-distribution in Eq.\,(\ref{app:eq:spec}) by a Breit-Wigner peak of width $\Gamma(T)$ resulting in
\begin{equation}
    \label{app:eq:mod_spec}
    \rho_{A_0}(\omega, T) = f_{\pi}^2m_{\pi}^2\frac{\Gamma(T)}{\pi}\frac{1}{2\omega_{\bold{0}}}\left(\frac{1}{(\omega-\omega_{\bold{0}})^2+\Gamma(T)^2}-\frac{1}{(\omega+\omega_{\bold{0}})^2+\Gamma(T)^2}\right)+\dots\,,
\end{equation}
where the second term is needed to ensure the antisymmetry of the spectral function in $\omega$\,\cite{Meyer:2011gj}. Expressing the correlator midpoint of the time-dependent Euclidean correlator $G_{A_0}(\beta/2,T)$ in terms of the spectral function $\rho_{A_0}$ with the help of Eq.\,(\ref{eq:axial_spec_func}) and using $\omega_{\bold{0}}=u_f m_{\pi}$ one can extract the pion velocity $u_f$ for different thermal pion widths $\Gamma(T)\in\{15,30,60\}$\,MeV. The results are shown in Table\,\ref{tab:thermal_width}.

\begin{table}[th]
\caption{Dependence of the pion velocity $u_f$ on a finite pion thermal width $\Gamma(T)$} 
\begin{tabular}{S[table-format=2]S[table-format=1.3(2)]}

\hline
\hline
{$\Gamma(T)\,[\text{MeV}]$} & {$u_f$} \\ 
\hline
15    & 0.783(19) \\
30    & 0.762(18) \\
60    & 0.671(21) \\
\hline
\hline
\end{tabular}
\label{tab:thermal_width}
\end{table}
We find that the extracted estimator of the pion velocity $u_f = 0.787(16)$ (assuming the presence of a discrete delta term in the spectral function) is consistent with a Breit-Wigner approach up to pion thermal widths $\Gamma(T) \approx 30\,\text{MeV}$.

\subsection{Spectral function reconstruction with the Backus-Gilbert method}
\label{sec:Backus}

\begin{figure}[tp]
	\includegraphics[scale=0.48]{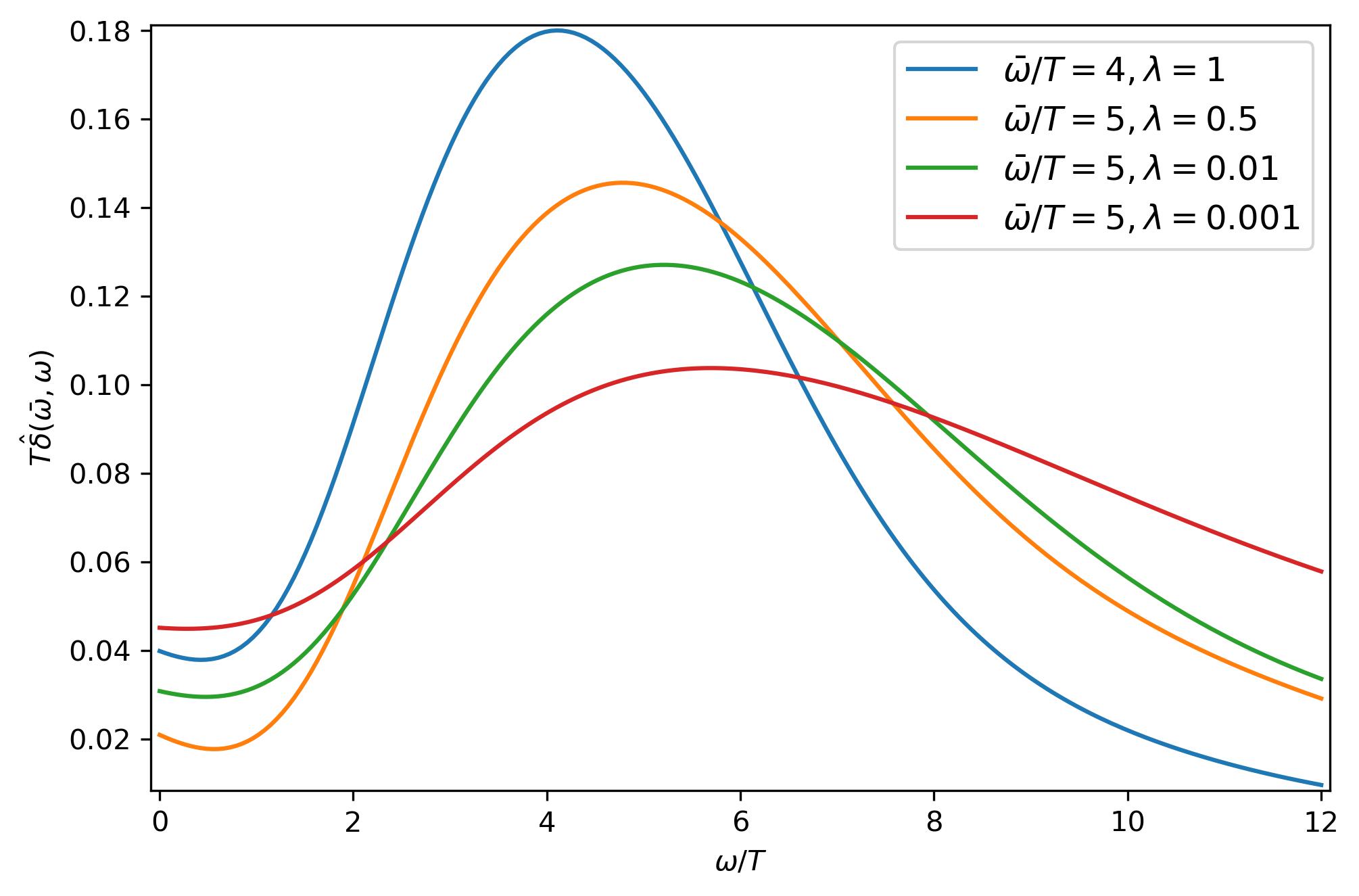}
	\includegraphics[scale=0.48]{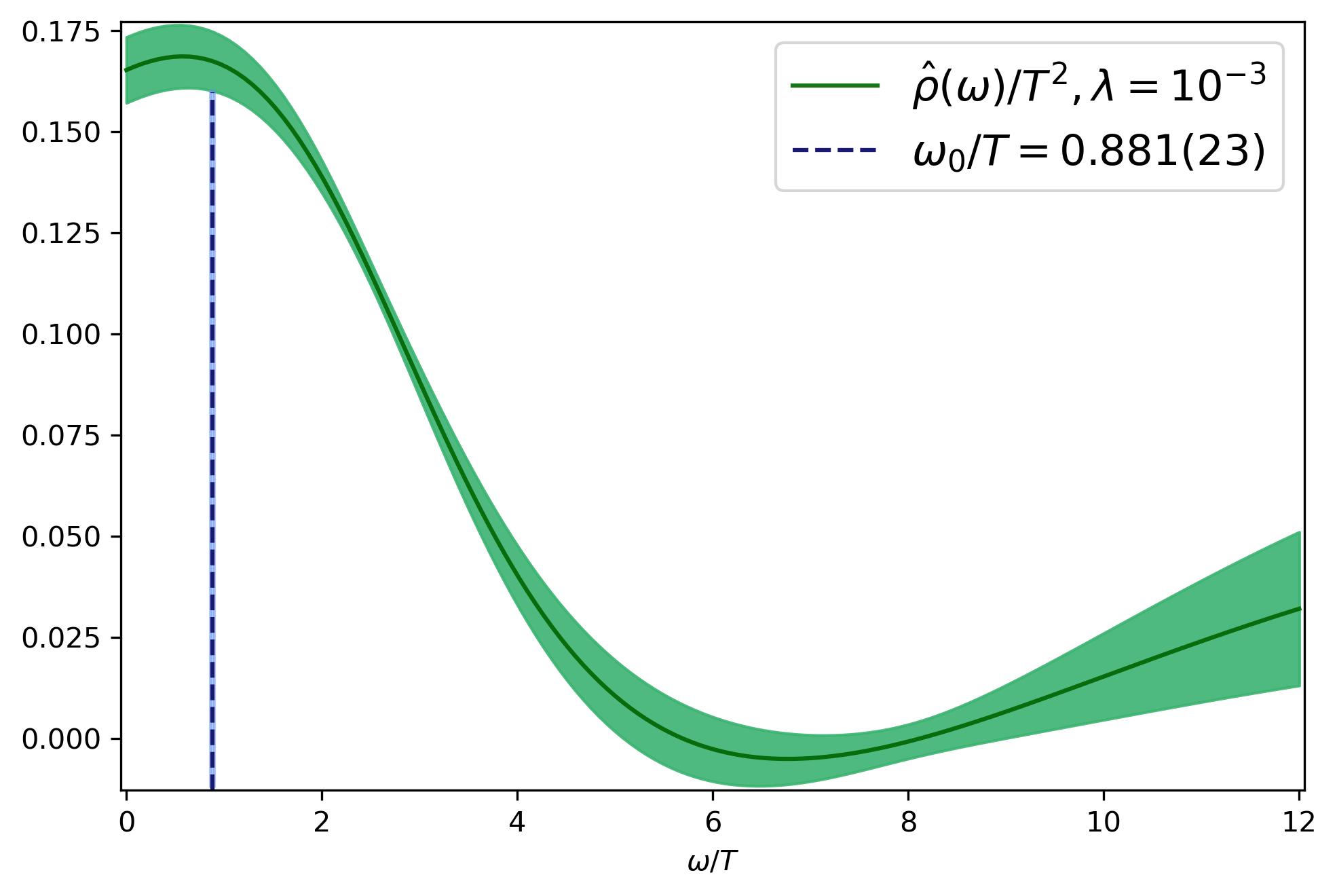}
\caption{
  {\bfseries Left panel}: Some examples of resolution functions for different values of $\lambda$, centered around $\bar{\omega}/T$. {\bfseries Right panel}: Estimator of the spectral function $\hat{\rho}_{A_0}(\omega,T)/T^2$. The blue dashed line corresponds to the location of the expected position of the pole $\omega_{\bold{0}}$ according to Eq.\,(\ref{eq:dispersion}). 
}
\label{fig:res_and_spec}
\end{figure}

In order to extract the spectral function $\rho_{A_0}(\omega)$ at zero momentum from the corresponding temporal Euclidean correlator, $G_{A_0}(\tau_i,T),\ \tau\equiv x_0$, one has to invert the analogue for $G_{A_0}$ of Eq.\,(\ref{eq:axial_spec_func}) with a kernel $K(\tau_i,\omega) = \text{cosh}(\omega(\beta/2-\tau_i))/\text{sinh}(\omega\beta/2)$, encountering a numerically ill-posed problem. 
A possible approach dealing with this task is the Backus-Gilbert method\,\cite{Backus:1968}. We adopt the notation of Ref.\,\cite{Brandt:2015sxa}, where the method has first been applied to lattice QCD. It should be emphasized that, with this approach, no particular ansatz needs to be made for the spectral function. The Backus-Gilbert method provides an estimator for the smeared axial spectral function
\begin{equation}
    \label{eq:smeared_sf}
    \hat{\rho}_{A_0}(\bar{\omega}) = \sum_{i=1}^{N_{\tau}} q_i(\bar{\omega})G_{A_0}(\tau_i) = \sum_{i=1}^{N_{\tau}} q_i(\bar{\omega})\int_0^{\infty}\mathrm{d}\omega\,K(\tau_i,\omega)\rho_{A_0}(\omega)\,,   
\end{equation}
built from the lattice correlator data $G_{A_0}(\tau_i)$. Note that the coefficients $q_i$ depend on some reference value $\bar{\omega}$ around which the so called resolution function (or averaging kernel),
\begin{equation}
    \label{eq:resolution_function}
    \hat{\delta}(\bar{\omega},\omega) = \sum_{i=1}^{N_{\tau}} q_i(\bar{\omega})K(\tau_i,\omega)\,,
\end{equation}
is concentrated. It is normalized according to 
\begin{equation}
    \label{eq:norm_cond}
    \int_0^{\infty}\mathrm{d}\omega\,\hat{\delta}(\bar{\omega},\omega)=1\,.
\end{equation}
Since the kernel $K(\tau_i,\omega)$ has a singularity in the limit $\omega\rightarrow 0$, it is advantageous to introduce a rescaling function 
\begin{equation}
    \label{eq:resc_func}
    f(w) = \text{tanh}(\omega \beta/2)\,,
\end{equation}
redefining the regularized kernel to be $K_f(\tau_i,\omega) = f(\omega)K(\tau_i,\omega)$.
This allows us to rewrite Eq.\,(\ref{eq:smeared_sf}) for the smeared and rescaled spectral function as
\begin{equation}
    \label{eq:smeared_sp_out_of_res_fun}
    \frac{\hat{\rho}_{A_0}(\bar{\omega})}{f(\omega)} = \int_0^{\infty}\mathrm{d}\omega\,\hat{\delta}(\bar{\omega},\omega)\frac{\rho_{A_0}(\omega)}{f(\omega)}\,.
\end{equation}
Inspecting Eq.\,(\ref{eq:smeared_sp_out_of_res_fun}), the desirable resolution function would be a Dirac delta distribution centered at $\bar{\omega}$. 
However, it has to satisfy Eq.\,(\ref{eq:resolution_function}) at the same time.
In order to make the resolution function as sharply centered around $\bar{\omega}$ as possible, we minimize the second moment of its square subject to
the constraint in Eq.\,(\ref{eq:norm_cond}). Therefore we minimize the following functional,
\begin{align}
    \label{eq:functional_to_minimize}
    \mathcal{F}[q_i(\bar{\omega})] & = \int_0^{\infty}\mathrm{d}\omega(\omega-\bar{\omega})^2\left[\hat{\delta}(\bar{\omega},\omega)\right]^2 - \alpha\left(\int_0^{\infty}\mathrm{d}\omega\,\hat{\delta}(\bar{\omega},\omega)-1\right) \notag \\
    &=\sum_{i,j=1}^{N_{\tau}} q_i(\bar{\omega})\left[\int_0^{\infty}\mathrm{d}\omega\,K_i(\tau_i,\omega)(\omega-\bar{\omega})^2K_j(\tau_j,\omega)\right]q_j(\omega) \ - \notag \\
     &\ \ \ \ \ -\sum_{i=1}^{N_{\tau}}\alpha\left(q_i(\bar{\omega})\int_0^{\infty}\mathrm{d}\omega\, K_i(\tau_i,\omega) -1 \right) \notag \\
     &\equiv \sum_{i,j=1}^{N_{\tau}} q_i(\bar{\omega})W_{ij}(\bar{\omega})q_j(\bar{\omega}) - \sum_{i=1}^{N_{\tau}} \alpha\left(q_i(\bar{\omega})R_i(\bar{\omega})-1\right)\,,
\end{align}
with $\alpha$ being a Lagrange multiplier. In practice the matrix $W_{ij}(\bar{\omega})$ is very close to being singular and needs a regularization procedure,
\begin{equation}
    \label{eq:regularization_W}
    W_{ij}(\bar{\omega})\rightarrow W_{ij}^{\text{reg.}}(\bar{\omega}) = \lambda W_{ij}(\bar{\omega}) + (1-\lambda)\text{Cov}[G_{A_0}]_{ij}\,,\quad\quad 0<\lambda<1\,.
\end{equation}
The resulting coefficients that minimize Eq.\,(\ref{eq:functional_to_minimize}) are then given by
\begin{equation}
    \label{eq:coeff_q}
    q_j(\bar{\omega}) = \frac{\sum_{k=1}^{{N_{\tau}}}\left(W^{\text{reg.}}(\bar{\omega})^{-1}\right)_{jk}R_k}{\sum_{i,l=1}^{N_{\tau}}R_i\left(W^{\text{reg.}}(\bar{\omega})^{-1}\right)_{il} R_l}\,.
\end{equation}
The values of $\lambda$ quoted below refer to units in which all dimensionful quantities are turned into dimensionless ones by appropriate powers of temperature.
Some examples for the resolution function $T \hat{\delta}(\bar{\omega},\omega)$ for different values of $\lambda$ are shown in the left panel of Fig.\,\ref{fig:res_and_spec}. The right panel of Fig.\,\ref{fig:res_and_spec} shows the smeared and rescaled  axial spectral function $\hat{\rho}_{A_0}(\bar{\omega},T)/T^2$ with $\lambda=10^{-3}$. It demonstrates model-independently that the axial-charge correlator is dominated by low frequencies. Furthermore, the predicted position of the quasiparticle mass $\omega_{\bold{0}}$ is close to
the peak of the smeared spectral function.

\subsection{Comparison with results from the literature}
\label{sec:comparison_lit}
Comparing our pion quasiparticle mass $\omega_{\bold{0}}(T)$ and quasiparticle decay constant $f_{\pi}^t(T)$ at $T=128$\,MeV to the matching quantities at the corresponding zero temperature ensemble we get $\omega_{\bold{0}}(T)/ m_{\pi}^0 = 0.880(25)$ and $f_{\pi}^t(T)/ f_{\pi}^0 = 1.039(26)$. Thus, the quasiparticle mass decreases at finite temperature while the quasiparticle decay constant increases. This behavior is similar to what is found in a ChPT calculation at two loops (see Ref.\,\cite{Toublan:1997rr}, Fig.\,3 and Fig.\,4). The reduction of the quasiparticle mass therein is $\approx 0.9$. On the other hand the quasiparticle decay constant increases by a factor of approximately 1.06.

Regarding the screening pion mass $m_{\pi}$, we found that it increases with temperature compared to $m_{\pi}^0$. The ratio is $m_{\pi}/m_{\pi}^0 = 1.125(27)$  This statement is also supported by the study of Son and Stephanov near the chiral phase transition\,\cite{Son:2001ff}. The screening mass that we obtained is larger than
what one would expect based on a short linear extrapolation to $T=128$\,MeV
using the continuum extrapolated data presented in a study with $N_{\mathrm{f}}=2+1$ highly-improved
staggered fermions in Ref.\,{\cite{Bazavov:2019www}}. However, we note that their study does not have any lattice data for the temperature that we used.

In the recent publication\,\cite{Goderidze:2022vlm}, the authors work out the pion damping width and the pion spectral function in the framework of a $SU(2)$ Nambu-Jona-Lasinio (NJL) model for a few temperatures below the critical temperature $T_c^{\text{NJL}}$ = 190\,MeV. They observe that the position of the peak of the pion spectral function at vanishing momentum $\bold{p}$ is moving to the right for increasing temperatures $T/T_c^{\text{NJL}} \in \{0,0.79,0.89,0.97\}$ (see Fig.\,(3) in \cite{Goderidze:2022vlm}). This contradicts our observation of the pion pole mass being reduced at finite temperature.

\section{Quark number susceptibility}
\label{sec:suscep}
With $\mathcal{N}_q = \int \mathrm{d}^3x\, V_0(0,\bold{x})$, $V_0$ being the time component of the local vector current\footnote{Note the additional normalization factor 1/2 resulting in an overall factor of 1/2 for the correlator and therefore also for the QNS.
}, the usual definition of the quark number susceptibility (QNS) for a given flavor is given by
\begin{align}
    \label{eq:QNS_1}
    \chi_q(T) = \left. \frac{\partial \rho_q}{\partial \mu_q} \right\vert_{\mu_q=0} \,.
\end{align} 
It measures the response of the quark number density,
\begin{align}
    \label{eq:quark_number_density}
    \rho_q = \frac{1}{V} \frac{\text{Tr}\left[\mathcal{N}_q\,e^{-\beta(\mathcal{H}-\mu_q\mathcal{N}_q)}\right]}{\text{Tr}\left[e^{-\beta(\mathcal{H}-\mu_q\mathcal{N}_q)}\right]} = \frac{\langle\mathcal{N}_q\rangle}{V}\, ,
\end{align}
to an infinitesimal change in the quark chemical potential $\mu_q \rightarrow \mu_q + \delta\mu_q$.

On the lattice, we define the quark number susceptibility as
\begin{align}
\label{QNS}
   \chi_q(x_0, T) =Z_V^2(g_0^2)\, \beta\, \int \mathrm{d}^3x\, \langle V_{0}^a(x_0,\bold{x})\,V_{0}^a(0,\bold{0}) \rangle\,, \quad \quad x_0 \neq 0\,.
\end{align}
Note, that for the QNS no improvement of the vector current is needed.

\begin{figure}[tp]
	\includegraphics[scale=0.75]{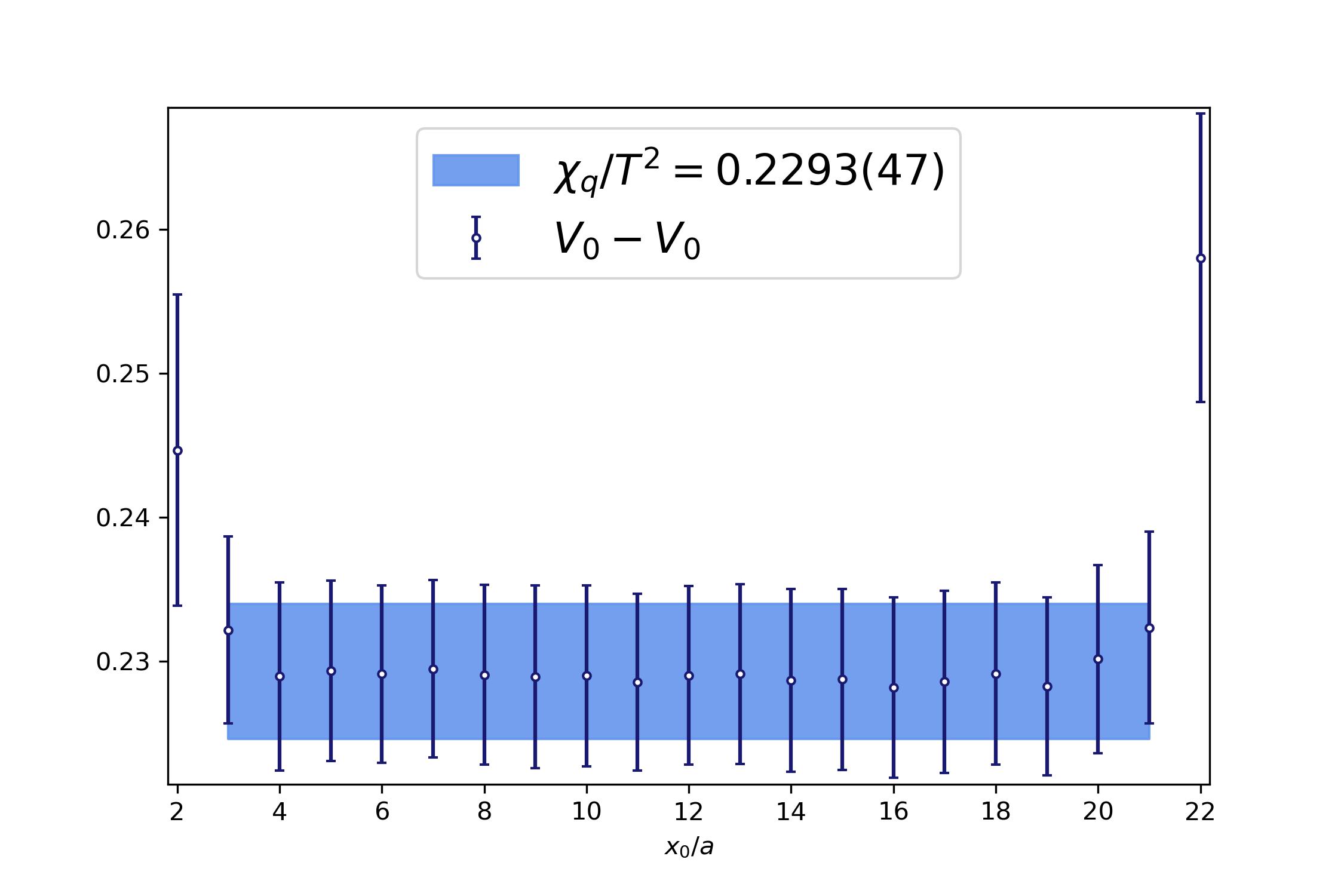}
	\caption{Quark number susceptibility extracted from the local vector current correlator, Eq.\,(\ref{QNS}). The mean and error have been obtained from a correlated fit in the range [3,21].
	\label{fig:QNS}}
\end{figure}
The result is shown in Fig.\,\ref{fig:QNS} in temperature units. Please note that we are not including any contributions from disconnected diagrams in our result for the QNS and in this approximation it is proportional to the isospin susceptibility. In Ref.\,\cite{Borsanyi:2011sw} (see Table\,I) the quark number susceptibility was determined as a function of the temperature using 2+1 dynamical staggered
quark flavors and, additionally, a continuum extrapolation was done. Taking into account the different normalization factor, their results are
$\chi_q(T)/T^2 = 0.216(46)$ and $\chi_q(T)/T^2 = 0.241(44)$ for the temperatures $T = 125$\,MeV and $T = 130$\,MeV, respectively (see Table\,I of Ref.\,\cite{Borsanyi:2011sw}).
Our result, $\chi_q(T)/T^2 = 0.2293(47)$ is compatible with both of these results. Although we did not perform a continuum extrapolation, our lattice spacing is around $2/3$ of the finest lattice spacing employed in Ref.\,\cite{Borsanyi:2011sw}, so beside the larger errors of the results from Ref.\,\cite{Borsanyi:2011sw}, the presence of only small cut-off effects may also explain the good agreement.
Next, we are going to compare our lattice estimate for the QNS with the hadron resonance gas (HRG) model and also test an alternative HRG employing our modified dispersion relation for the pion quasiparticle.

\subsection{Comparison with the hadron resonance gas model}
\label{subsec:HRG}

The HRG model\,\cite{Hagedorn:1984hz, Karsch:2003vd} describes the thermodynamic properties and the quark number
susceptibilities of the low-temperature phase rather well. It assumes that the thermodynamic properties of the system are given by the sum of the partial contributions of non-interacting hadron species, i.e.
\begin{align}
    \label{eq:HRG partition function}
    \text{ln}[Z(T,V)] = -\frac{V}{2\pi^2}\sum_i\int_0^{\infty}\,\mathrm{d}p\, p^2\,\text{ln}[1-\eta_i\,e^{-\sqrt{m_i^2+\bold{p}^2}/T}]\,,
\end{align}
where $\eta_i = \pm 1$ takes into account bosons and baryons respectively. The sum extends over all resonances up to a mass of 2.0\,GeV, since for most of them the width is not large compared to the temperature.

The quark number susceptibility can be obtained as the sum\,\cite{Brandt:2015aqk}
\begin{equation}
    \chi_q(T) = (\chi_q)_{\text{mesons}} + (\chi_q)_{\text{baryons}}\,,
\end{equation}
where
\begin{align}
    \label{eq:QNS HRG mesons}
    \frac{(\chi_q)_{\text{mesons}}}{T^2} &= \frac{2\beta^3}{3}\sum_{\text{multiplets}}(2J+1)I(I+1)(2I+1)\int\frac{\mathrm{d^3}\bold{p}}{(2\pi)^3}\,f_{\bold{p}}^B(1+f_{\bold{p}}^B)\,,\\
    \label{eq:QNS HRG baryons}
    \frac{(\chi_q)_{\text{baryons}}}{T^2} &= \frac{2\beta^3}{3}\sum_{\text{multiplets}}(2J+1)I(I+1)(2I+1)\int\frac{\mathrm{d^3}\bold{p}}{(2\pi)^3}\,f_{\bold{p}}^F(1-f_{\bold{p}}^F)\,,
\end{align}
and $f_{\bold{p}}^{B/F} = 1/[e^{\beta\omega_{\bold{p}}}\mp 1]$ are the Bose-Einstein and Fermi-Dirac distributions.
The sums are carried out over all multiplets of spin $J$ and isospin $I$ that are not identical. Especially particles and antiparticles have to be considered separately. This results in an additional factor of two in the baryon case and for mesons with strange quark constituents.\\
An alternative to the HRG model is to only include the
pion contribution, however taking into account the modified dispersion relation (\ref{eq:dispersion}) at low momenta,
\begin{align}
    \label{eq:QNS with modified dispersion relation}
    \frac{\chi_q}{T^2} = 4\beta^3\int_{\abs{\bold{p}}<\Lambda_p}\frac{\mathrm{d^3}\bold{p}}{(2\pi)^3}\,f_{\bold{p}}^B(\omega_{\bold{p}})(1+f_{\bold{p}}^B(\omega_{\bold{p}}))\,, 
\end{align}
where $\Lambda_p = 400\,\text{MeV}$  is about the momentum scale at which the predictions of the thermal chiral effective theory were seen to break down in Ref.\,\cite{Brandt:2015sxa}. Note that in this model the sum over the resonances is absent. The contributions of the other hadrons are taken into account indirectly via the modified dispersion relation, since the collisions of the pions among themselves and with other hadrons give rise to the modified pion dispersion relation.
Employing Eqs.\,(\ref{eq:QNS HRG mesons}-\ref{eq:QNS HRG baryons}) within the HRG model, summing all resonances up to a mass of 2\,GeV, one obtains $\chi_q(T)/T^2 = 0.2428$ which is $5.8\%$ above our lattice estimate $\chi_q(T)/T^2 = 0.2293(47)$.

In the HRG model the pion contributes $\left.\chi_q/T^2\right\vert_{\text{pion}}=0.1890$ corresponding to $77.9\%$ of the total QNS followed by a contribution of $15.4\%$ of the vector and pseudoscalar meson octets ($7.3\%$ of this is attributable to the $\rho$ vector meson). The baryon octet and decuplet contributes $3.8\%$, the largest portion ($2.5\%$) stemming from the $\Delta$ resonance due to the large spin degeneracy factor. Heavier meson and baryon resonances up to a mass of 2\,GeV contribute the remaining $2.9\%$ to the final result. It is questionable whether resonances whose full width is higher than the temperature should be taken into account. For instance the $K_0^*(700)$ resonance has a full Breit-Wigner width $(478\pm50\,\text{MeV})$ of nearly four times the temperature and has therefore been neglected.

Making use of Eq.\,(\ref{eq:QNS with modified dispersion relation}) with $u_m=0.786$ and a screening pion mass $m_{\pi} = 144\, \text{MeV}$ one obtains $\chi_q(T)/T^2 = 0.2163$ which is $5.3\%$ below the lattice estimate. At this point we have only integrated up to the momentum cut off $\Lambda_p = 400\,\text{MeV}$ since it is not clear if the thermal width of the pion is still negligible for $\abs{\bold{p}}>\Lambda_p$ and, as a consequence, including contributions from higher momenta may not be justified. However, this model is not very predictive as it depends very strongly on the choice of the momentum cutoff.
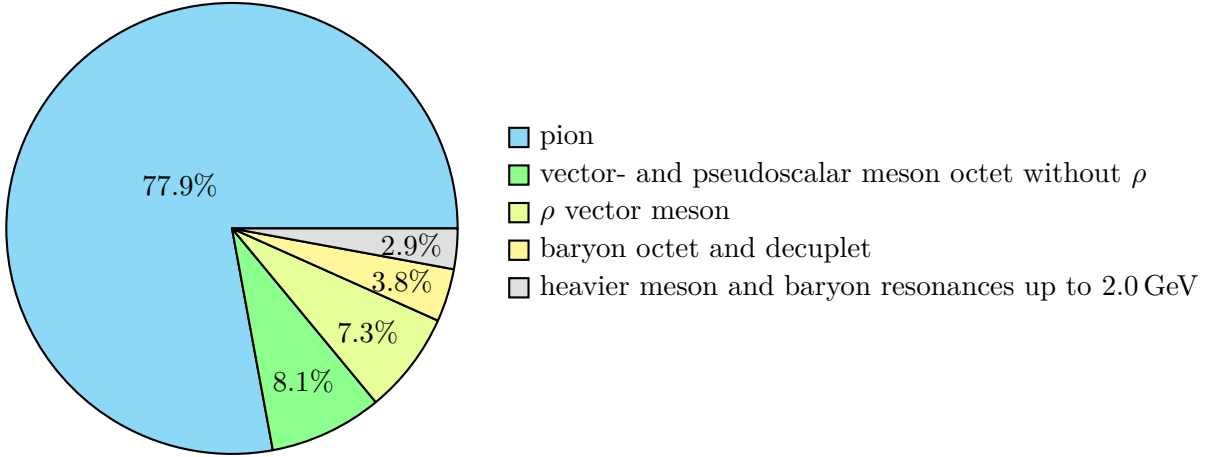
\begin{figure}[tp]
\begin{tikzpicture}
 \pie[
    color = {
        cyan!=40!, 
        green!45!, 
        lime!40, 
        yellow!50,
        lightgray!50},
        text = legend 
        ]
        {77.9/pion,
    8.1/vector- and pseudoscalar meson octet without $\rho$,
    7.3/ $\rho$ vector meson,
    3.8/baryon octet and decuplet,
    2.9/heavier meson and baryon resonances up to $2.0\,\text{GeV}$
    }
 \end{tikzpicture}
 \caption{Relative composition of the total quark number susceptibility predicted by the hadron resonance gas model.
	\label{fig:pie diagram}}
\end{figure}

\section{Order parameters for chiral symmetry restoration}
\label{sec:order_param}

In this section, several order parameters for chiral symmetry restoration are investigated. Based on the screening pion quantities $m_{\pi}^2$ and $f_{\pi}^2$ presented in Sec.\,\ref{sec:results}, we first evaluate an ‘effective chiral condensate’ based on the Gell-Mann--Oakes--Renner relation.
Additionally, we explore two Euclidean-time dependent thermal correlation functions that are order parameters for chiral symmetry
and compare them to their zero-temperature counterparts.
We begin with the $(PA_0)$-correlator, which contains the pion pole that we have studied in Sec.\,\ref{sec:results}.
We then consider the difference of the (isovector) vector and axial-vector correlators.
In the QCD vacuum, the corresponding spectral functions are measured experimentally in $\tau$ decays~\cite{Davier:2005xq}.
They become degenerate in the chirally restored phase of QCD. Their temperature dependence in the chirally broken phase
has been studied extensively in the framework of hadronic models supplemented by sum rules~\cite{Rapp:1999ej,Kapusta:1993hq,Hohler:2013eba}.

\subsection{The Gell-Mann--Oakes--Renner (GOR) relation}
\label{sec:GOR}

Following Ref.\,\cite{Brandt:2014qqa}, we introduce a ‘effective chiral condensate’ based on the Gell-Mann--Oakes--Renner (GOR) relation,
\begin{equation}
    \langle\bar{\psi}\psi\rangle_{\text{GOR}}^{\text{GF}}
     \equiv -\frac{f_{\pi}^2m_{\pi}^2}{m_{\text{q}}}\,. 
\end{equation}
For $m_{\text{q}} \rightarrow 0$ it matches the actual chiral condensate. Additionally, since above $T_c$, $m_{\pi} \sim T$ and $f_{\pi} \sim m_{\text{q}}$, $\langle\bar{\psi}\psi\rangle_{\text{GOR}}^{\text{GF}}$ is of $\mathcal{O}(m_{\text{q}}\,T^2)$. Thus, it serves as an order parameter for chiral symmetry. Using $m_{\text{q}} = m_{\text{PCAC}}$ and the screening quantities of Table\,\ref{tab:results} we obtain
\begin{equation}
  \left|\langle\bar{\psi}\psi\rangle_{\text{GOR}}^{\text{GF}}\right|^{1/3} = 286(5)\, \text{MeV}\,. 
\end{equation}
The value of the chiral condensate has been extracted in the gradient flow scheme just like the PCAC mass (see Sec.\,\ref{sec:lat_impl}).
Comparing with the chiral condensate on the corresponding zero-temperature ensemble \cite{Ce:2022kxy}, we get
\begin{equation}
    \label{eq:reduction_quark_cond}
    \left[\frac{\langle\bar{\psi}\psi\rangle_{T\approx 128\,\text{MeV}}}{\langle\bar{\psi}\psi\rangle_{T\approx 0\,\text{MeV}}}\right]_{\text{GOR}} \equiv \frac{(f_{\pi}^2m_{\pi}^2)_{T\approx 128\,\text{MeV}}}{(f_{\pi}^2m_{\pi}^2)_{T\approx 0\,\text{MeV}}} = 0.84(5)\,,
\end{equation}
which corresponds to a reduction by 16\%. This reduction is compatible within the scope of the error with a three-loop result of Gerber and Leutwyler (see Ref.\,\cite{Gerber:1988tt}, Fig.\,5). 

\subsection{The \texorpdfstring{$(PA_0)$}{(PA0)}-correlator}

In Ref.\,\cite{Brandt:2014qqa} it was shown that the  temporal $(PA_0$)-correlator can be predicted exactly in the chiral limit,
\begin{equation}
  \label{eq:PA0_corr}
        G_{PA_0}(x_0,T) = \frac{\langle\bar{\psi}\psi\rangle}{2\beta}    \left(x_0-\frac{\beta}{2}\right)\,.
\end{equation}
As can be seen from Eq.\,(\ref{eq:PA0_corr}) the $(PA_0)$-correlator is antisymmetric around $\beta/2$. Consequently, we set the point $x_0=12a$ to zero.
Since this correlator is proportional to the chiral condensate $\langle\bar{\psi}\psi\rangle$, it can serve as an order parameter for chiral symmetry restoration as well. Looking at the ratio of the thermal over reconstructed correlator [see Eq.\,(\ref{eq:reconstr})], we observe a more pronounced reduction by a factor of $\approx 0.80(7)$ compared to the reduction by a factor of $0.84(5)$ that we had estimated using the Gell-Mann–Oakes–Renner reation [see Eq.\,(\ref{eq:reduction_quark_cond}]. 

\begin{figure}[tp]
	\includegraphics[scale=0.75]{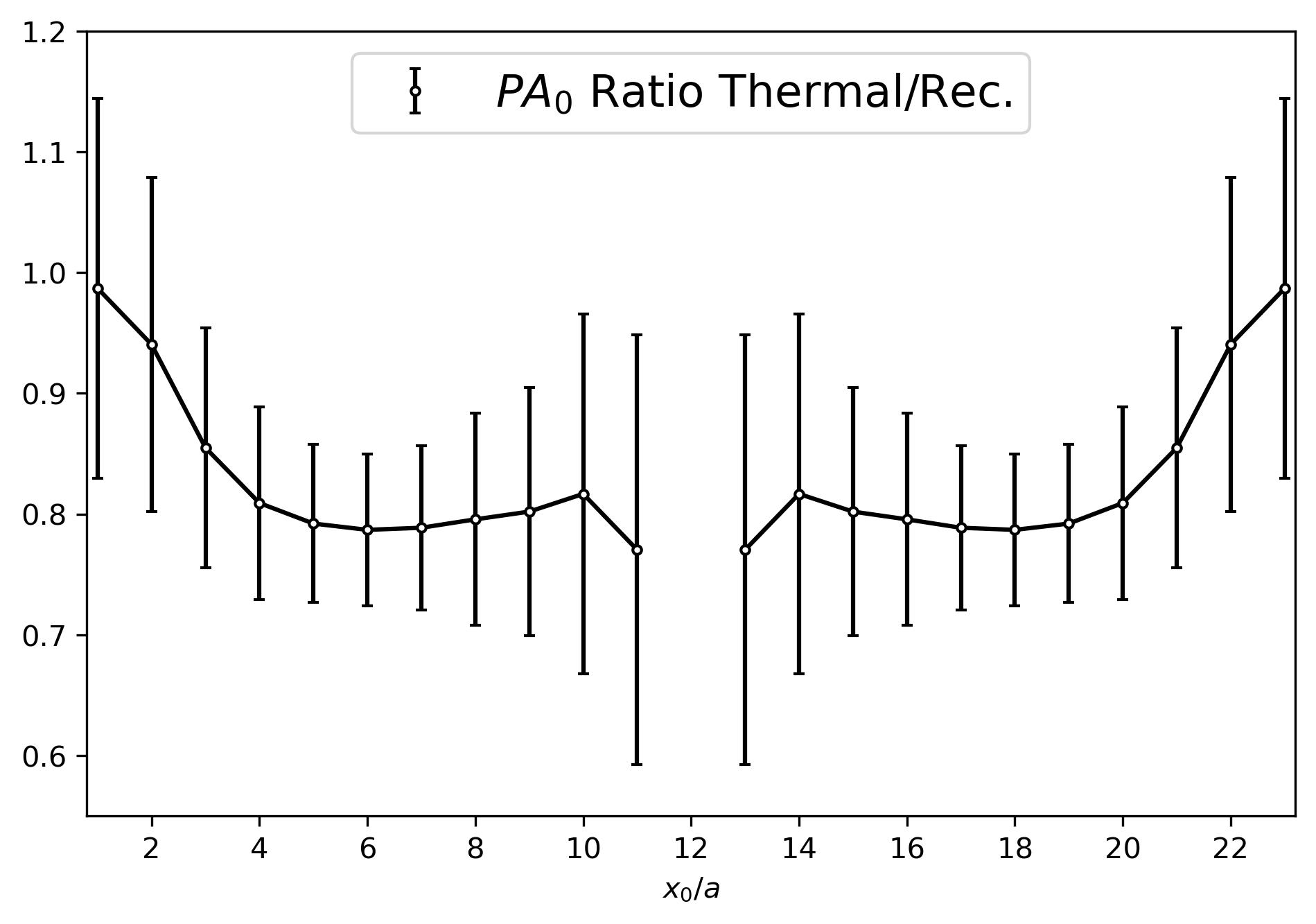}
	\caption{Ratio of the temporal thermal $(PA_0)$-correlator and  the reconstructed correlator $(PA_0)$-correlator.
	\label{fig:ratio_A0P}}
\end{figure}


\begin{figure}[tp]
	\includegraphics[scale=0.58]{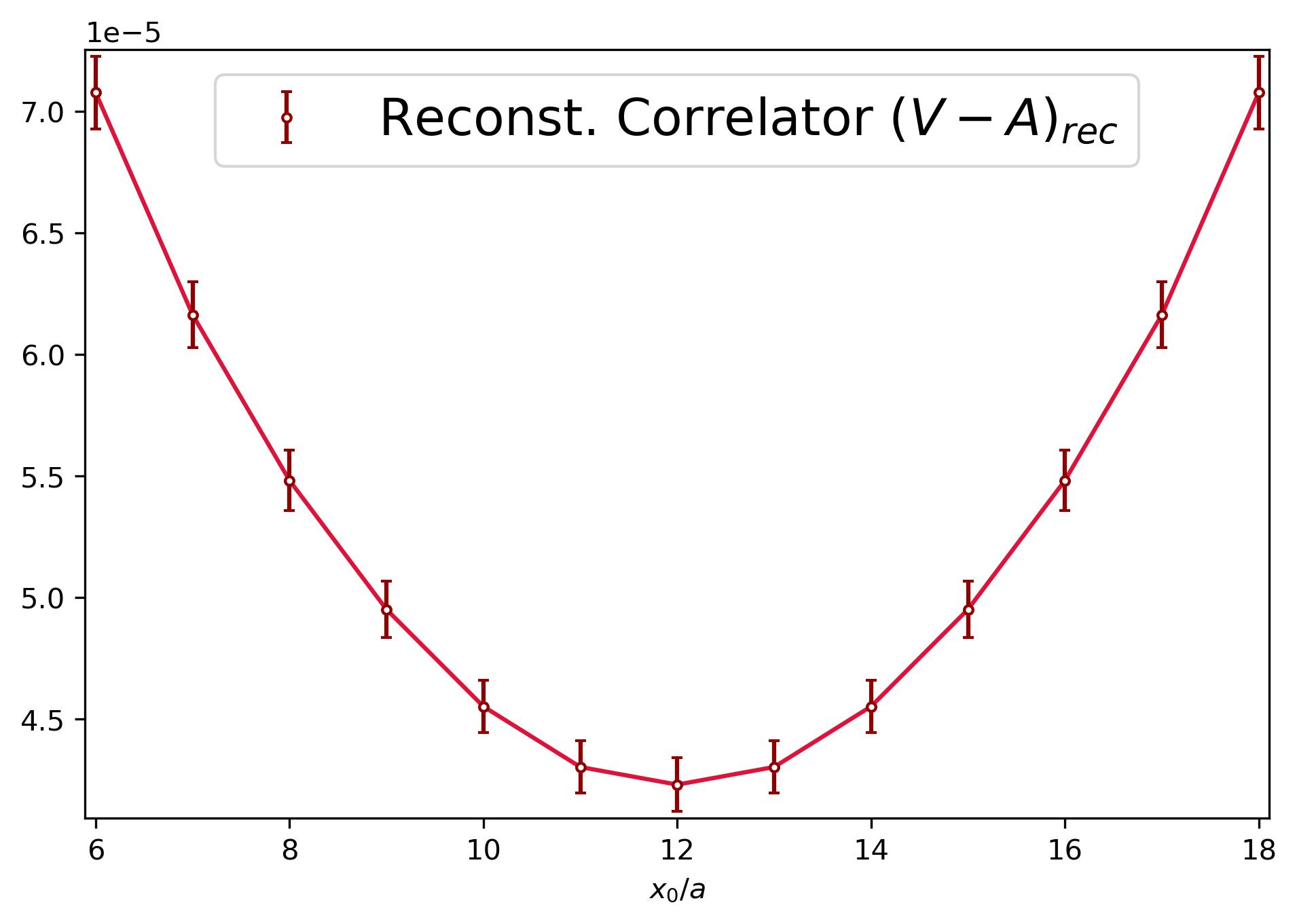}
	\includegraphics[scale=0.58]{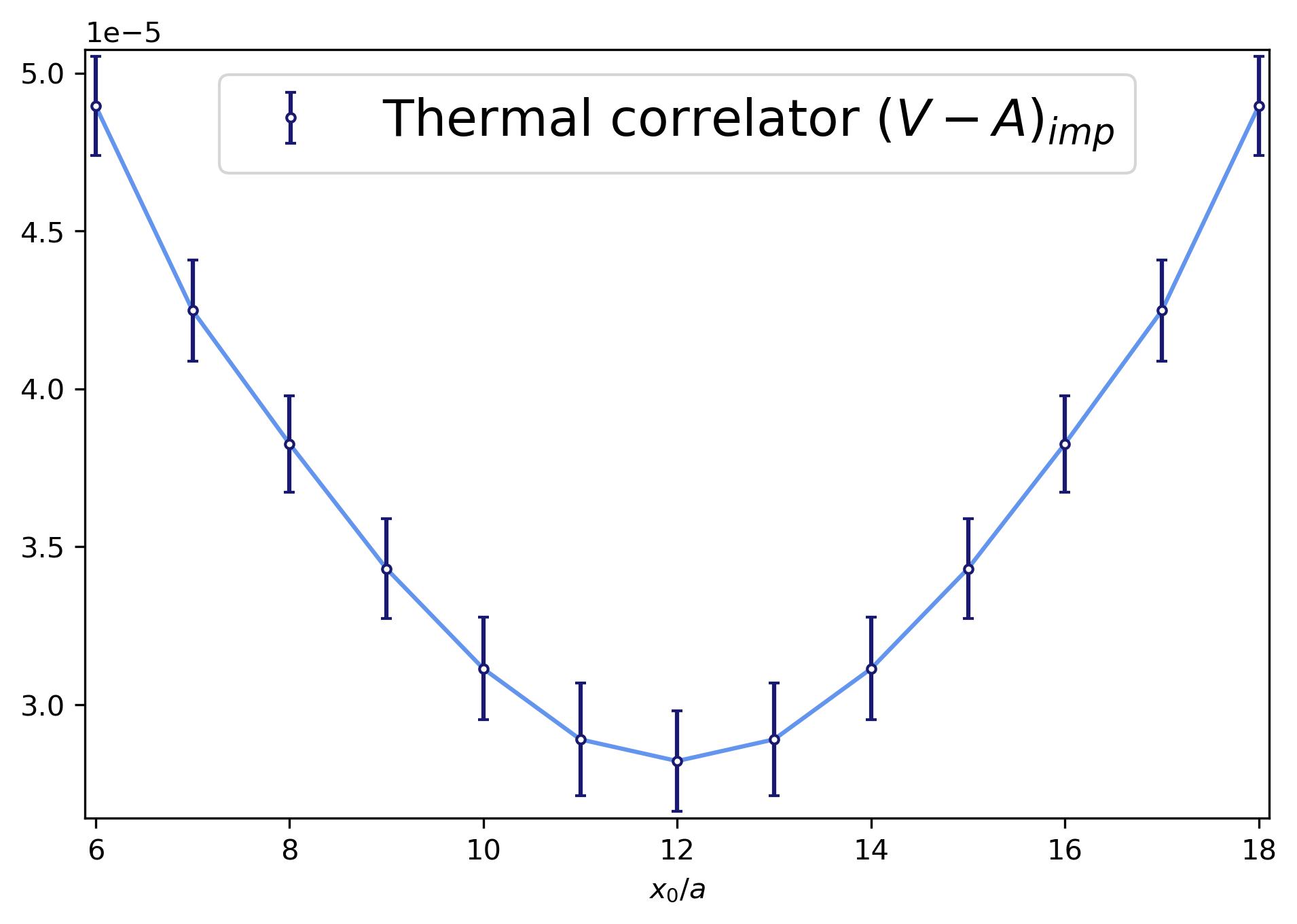}
	\includegraphics[scale=0.58]{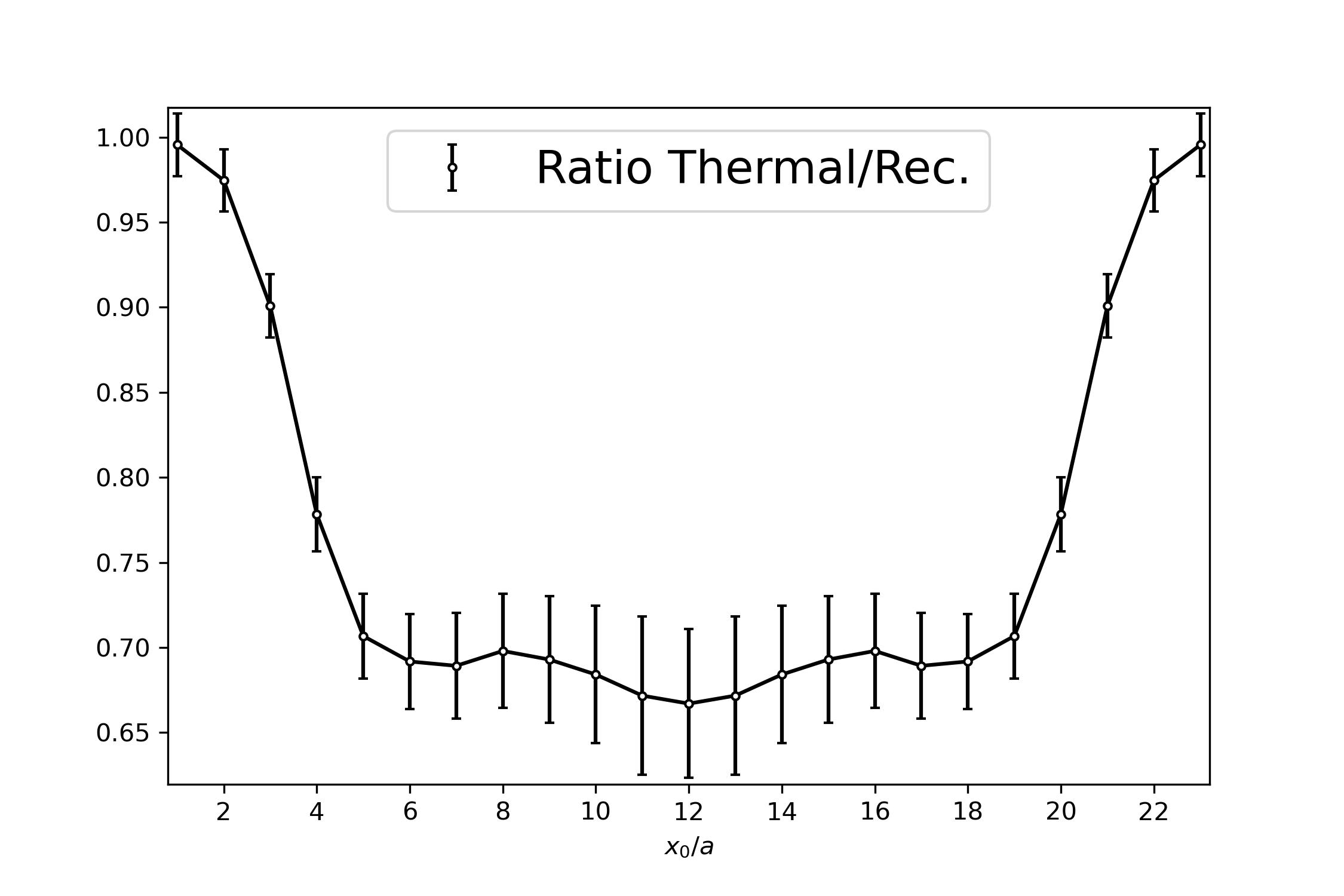}
\caption{
{\bfseries Top panel}: The reconstructed correlator for the difference ‘$V-A$’. {\bfseries Middle panel}: The difference of ‘$V-A$’ at $T \approx 128\, \text{MeV}$. {\bfseries Bottom panel}: Ratio of the difference ‘$V-A$’ and the difference of the reconstructed correlator ‘$(V-A)_{\text{rec}}$’.
}
\label{fig:V-A}
\end{figure}

\subsection{Dey-Eletsky-Ioffe mixing theorem at finite quark mass}
\label{sec:Ioffe}

Let us consider the following real-time correlators at finite temperature $T$:
\begin{align}
    \label{eq:corr}
    C_{\mu\nu}^{J,a,b}(q,T) = \frac{i\delta^{ab}\int \md^4x\,e^{iqx}\sum_n\bra{n}\mathbb{T}\{J_{\mu}^a(x)J_{\nu}^b(0)\}e^{-H/T}\ket{n}}{\sum_n\bra{n}e^{-H/T}\ket{n}} \ \ \ \ \ \ \  (J\in\{V,A\})\, ,
\end{align}
where the sum is over the full set of the eigenstates of the Hamiltonian $H$ and $\{a,b\}$ are isospin indices. To order $T^2$ it is sufficient to account only for the contributions of the two lowest states in Eq.\,(\ref{eq:corr}) – vacuum and one pion state. In Refs.\,\cite{Dey:1990ba,Eletsky:1992ay,Eletsky:1994rp} it was demonstrated, using PCAC current algebra, that the finite-temperature vector and axial-vector correlators can be described with the help of their vacuum counterparts. In terms of the corresponding spectral functions this statement reads
\begin{align}
    \label{eq:mixing_spectral_V}
    \rho_V(\omega,\bold{p},T) &= (1-\epsilon) \rho_V(\omega,\bold{p},T=0) + \epsilon     \rho_A(\omega,\bold{p},T=0)\, , \\
    \label{eq:mixing_spectral_A}
    \rho_A(\omega,\bold{p},T) &= (1-\epsilon) \rho_A(\omega,\bold{p},T=0) + \epsilon     \rho_V(\omega,\bold{p},T=0)\, ,
\end{align}
where $\epsilon \equiv T^2/(6(f_{\pi}^0)^2)$ is a temperature dependent expansion parameter in powers of the pion density. Notice that as a consequence of Eqs.\,(\ref{eq:mixing_spectral_V},\ref{eq:mixing_spectral_A}) the sum of the vector and axial-vector spectral function does not change when the temperature is switched on. Furthermore, the difference is proportional to its zero-temperature equivalent:
\begin{align}
    \label{eq:diff_spectral}
    \rho_V(\omega,\bold{p},T)-\rho_A(\omega,\bold{p},T)=(1-2\epsilon)\left[\rho_V(\omega,\bold{p},T=0)-\rho_A(\omega,\bold{p},T=0)\right]\, .
\end{align}
As a consequence the above quantity serves as an order parameter for chiral symmetry restoration. Thus, in the following we will investigate its behavior even for non-zero quark mass. To do so, we consider the difference ‘$V-A$’ of the corresponding $\mathcal{O}(a)$-improved temporal correlators projected to zero momentum
\begin{align}
    \label{eq:diff_corr}
    \delta^{ab}\left[G_V(x_0,T,\bold{p}=0)-G_A(x_0,T,\bold{p}=0)\right] \equiv -\frac{1}{3}\int \md^3x\, \sum_{i=1}^3\left[\langle V_i^a(x)V_i^b(0)\rangle - \langle A_i^a(x)A_i^b(0)\rangle\right]\, .
\end{align}

 In order to obtain a comparable effectively zero-temperature quantity, we use the corresponding quasi zero-temperature E250 ensemble of size $192\times96^3$. This is achieved by calculating the ‘reconstructed’ correlator $G_V^{\text{rec}}-G_A^{\text{rec}}$ for the difference, i.e., the thermal Euclidean correlator that would be realized if the spectral function was unaffected by thermal effects. Following a method first proposed in Ref.\,\cite{Meyer:2010}, we define our reconstructed correlators as
\begin{align}
    \label{eq:reconstr}
    G_J^{\text{rec}}(x_0,T,\bold{p}) = \sum_{m\in \mathrm{Z}}\,G_J(\abs{x_0+m\beta},0,\bold{p}) \ \ \ \ \ \ \ (J \in \{V,A\}). 
\end{align}
It is based on the identity of the kernel function 
\begin{align}
    \frac{\text{cosh}(\omega(\beta/2-x_0))}{\text{sinh}(\omega \beta/2)} = \sum_{m \in \mathrm{Z}}\, e^{-\omega\abs{x_0+m\beta}}\, .
\end{align}
The top and middle panel of Fig.\,\ref{fig:V-A} shows the difference ‘$V-A$’ for the thermal ensemble and the same quantity for the reconstructed correlator. Their ratio is shown in the bottom panel of Fig.\,\ref{fig:V-A}. For vanishing quark masses one would expect it to be flat consistent with Eq.\,(\ref{eq:diff_spectral}) obtained in the chiral limit. However, since chiral symmetry restoration is a long-distance effect, one expects that for physical quark masses the suppression of the $(V-A)$ spectral function happens mostly at low energies which translates to the longest (Euclidean) time accessible. This is consistent with the dip that we observe around the midpoint $(x_0=12)$ in Fig.\,\ref{fig:V-A}. Furthermore, the difference ‘$V-A$’ shows a significant reduction, by a factor of approximately 0.67 at $T \approx 128\,\text{MeV}$. Therefore, chiral symmetry restoration is already at an advanced stage in the spectral function.

\section{Conclusion}
\label{sec:conclusion}

In this work we have found that the zero-temperature pion mass `splits’ into a lower pion
quasiparticle mass $\omega_{\bold{0}}$ and a higher pion screening mass $m_{\pi}$ at finite temperature,
confirming the findings of Ref.\,\cite{Brandt:2015sxa} in QCD with two quark flavours ($u,d$).
Our results are also in good quantitative agreement with existing predictions of chiral perturbation theory:
see Secs.\,\ref{sec:comparison_lit} and~\ref{sec:GOR}.
Additionally, we have computed the two temperature-dependent parameters which determine the modified dispersion relation of the pion quasiparticle in the low-temperature phase of QCD [see Eq.\,(\ref{eq:dispersion})]. An assumption in determining the pion-velocity parameter $u$ was a discrete peak structure in the axial spectral function. Using instead a Breit-Wigner ansatz with a finite thermal width, we could confirm that -- within the statistical error -- our estimator of the pion velocity $u_f$ is stable within about three percent up to a finite pion width $\Gamma(T) \approx 30$\,MeV.
We have further employed the Backus-Gilbert method to show, independently of any model, that the axial correlator is indeed dominated by low frequencies.

The quark number susceptibility computed on the lattice has been compared to the predictions of the hadron resonance gas model as well as to the estimate where only pions are taken into account, however using their modified dispersion relation. The lattice estimate is found to lie approximately in the middle between these two predictions. Nonetheless, one should keep in mind the strong dependence on the momentum cutoff of the last approach. Hence, an analysis at non-vanishing momentum with a high resolution would be desirable in order to narrow down the validity of the chiral effective theory and, as a consequence, determine an appropriate value of the momentum cutoff $\Lambda_p$ more precisely.

Thirdly, we have investigated the degree of restoration of chiral
symmetry in two different channels, namely the pseudoscalar one and in
the difference of the (isovector) vector and axial-vector correlators (V-A).
We have done this by forming the ratio of the thermal correlator to the one reconstructed from the zero-temperature simulation.
Quantitatively, we found the (V-A) channel to exhibit a higher degree of chiral symmetry restoration.

Looking ahead, one might further ask if the relatively strong change
in the pion screening quantities is due to its Goldstone-boson nature
or if non-Goldstone hadrons are similarly modified by thermal effects~\cite{Aarts:2017rrl}.
We are  generating a thermal ensemble with $N_{\tau} =20$ and otherwise identical parameters. This choice corresponds to a
temperature $T= 154$\,MeV, right within the crossover regime.


\acknowledgments{This work was supported by the European Research Council (ERC) under the European 
Union’s Horizon 2020 research and innovation program through Grant Agreement
No.\ 771971-SIMDAMA, as well as by the Deutsche Forschungsgemeinschaft 
(DFG, German Research Foundation) through the Cluster of Excellence “Precision Physics,
Fundamental Interactions and Structure of Matter” (PRISMA+ EXC 2118/1) funded by
the DFG within the German Excellence strategy (Project ID 39083149).
T.H. is supported by UK STFC CG ST/P000630/1.
The generation of gauge configurations as well as the computation of correlators was 
performed on the Clover and Himster2 platforms at Helmholtz-Institut Mainz and on Mogon II 
at Johannes Gutenberg University Mainz.
We made use of the following libraries: GNU Scientific Library\,\cite{Galassi:2019},
GNU MPFR\,\cite{10.1145/1236463.1236468} and GNU MP\,\cite{Granlund:2020}.
Ardit Krasniqi wants to thank Renwick J. Hudspith for proofreading early versions of the draft.}

\clearpage
\appendix

\section{Renormalization process}
\label{app:ren}

Following \cite{Korcyl:2016ugy}, we renormalize the correlators in the following way:
\begin{align}
    \label{eq:renorm_V}
    G_V^{\text{ren.}} &= Z_V^2(g_0^2)(1+2am_qb_V(g_0^2))G_V\, ,\\
    \label{eq:renorm_A}
    G_A^{\text{ren.}} &= Z_A^2(g_0^2)(1+2am_qb_A(g_0^2))G_A\, ,\\
    G_{A_0}^{\text{ren.}} &= Z_A^2(g_0^2)(1+2am_qb_A(g_0^2))G_{A_0}\, ,\\
        G_{PA_0}^{\text{ren.}} &= Z_P(g_0^2) Z_A(g_0^2)(1+am_qb_A(g_0^2))G_{PA_0}\, ,\\
    \label{eq:renorm_P}
    G_P^{\text{ren.}} &= Z_P^2(g_0^2)G_P\, ,
    \end{align}
    
with $g_0^2 = 6/\beta$ being the bare gauge coupling and 
\begin{align}
    \label{eq:m_q}
    am_q = \frac{1}{2}\left(\frac{1}{\kappa_{\text{l}}}-\frac{1}{\kappa_{\text{cr.}}}\right)\, ,
\end{align}
being the bare subtracted quark mass.
The values for the renormalization constants $Z_J$ and the finite quark mass parameters $b_J$ are given in Table\,\ref{tab:ren_par}.

\begin{table}[tb]
\caption{Summary of the renormalization parameters.} 
\vspace{-0.4cm}
\begin{tabular}{lcS[table-format=1.9(2)]}
\\
\hline
\hline
$Z_V(g_0^2)$          & \cite{Gerardin:2018kpy}   & 0.73453(6) \\ 
$Z_A(g_0^2)$          & \cite{DallaBrida:2018tpn} & 0.76900(42)\\ 
$Z_P(g_0^2)$          & \cite{Campos:2018ahf}     & 0.34768 \\ \hline
$b_V(g_0^2)$          & \cite{Gerardin:2018kpy}   & 1.551(10) \\
$b_A(g_0^2)$          & \cite{Korcyl:2016ugy}     & 1.38(6) \\ 
\hline
$\kappa_{\text{cr.}}$ & \cite{Gerardin:2018kpy}   & 0.1371726(13) \\
$\kappa_{\text{l}}$   & \cite{Ce:2022eix}         & 0.137232867 \\
\hline
\hline
\end{tabular}
\label{tab:ren_par}
\end{table}

\section{Extracting \texorpdfstring{$f_{\pi}$}{fπ} out of \texorpdfstring{$A_1$}{A1}}
\label{app:extracting}

Let us denote the correlation function of a single state propagating forward as
\begin{align}
    \label{eq:forward}
    c_f(x_3) = c_f^0\, e^{-m_1x_3}\, ,
\end{align}
and analogously we denote the backward contribution as
\begin{align}
    \label{eq:backward}
    c_b(x_3) = c_b^0\, e^{-m_1(L-x_3)}\, .
\end{align}
Including the (tiny) contributions warping around the lattice the forward contribution becomes
\begin{align}
    c_f(x_3) &= c_f^0\left(e^{-m_1x_3} + e^{-m_1(L+x_3)}+\dots\right) \notag \\
           &= c_f^0\, e^{-m_1x_3}\sum_{n=0}^{\infty}\, e^{-nm_1L} \notag \\
           &= c_f^0\, e^{-m_1x_3} \frac{1}{1-e^{-m_1L}}\, .
\end{align}
Combining forward and backward contribution and comparing with Eq.\,(\ref{eq:axial_fit}) we obtain
\begin{align}
     G_A^s(x_3,T) = \frac{A_1^2 m_1}{2}\, \text{cosh}[(m_1(x_3-L/2)] = \frac{c_0}{1-e^{-m_1L}}\left(e^{-m_1x_3}+ e^{-m_1(L-x_3)}\right)\, .
\end{align}
Pulling out a factor of $e^{-m_1L/2}$ and reading  off $c_0 = \frac{1}{2}f_{\pi}^2m_1$ from Eq.\,(\ref{eq:asymp}) we can finally link the screening pion decay constant $f_{\pi}$ with the fit parameter $A_1$ as follows,
\begin{align}
    \label{eq:decay_const}
    f_{\pi} =A_1 \sqrt{ \text{sinh}\left(m_{1}L/2\right)}
\end{align}

\section{Chiral effective theory Lagrangian of Son and Stephanov}
\label{app:Son}
In the chiral effective theory approach of Son and Stephanov \cite{Son:2001ff}\cite{Son:2002ci} the dynamics of the pions at finite temerpature is described by the Lagrangian,
\begin{align}
    \label{eq:effective Lagrangian}
    \mathcal{L}_{\text{eff}} = \frac{f_t^2}{4}\langle\nabla_0\Sigma\nabla_0\Sigma^{\dagger}\rangle - \frac{f_{\pi}^2}{4}\langle\partial_i\Sigma\partial_i\Sigma^{\dagger}\rangle + \frac{m_{\pi}^2f_{\pi}^2}{2}\text{Re}\langle\Sigma\rangle\,,
\end{align}
where $\Sigma$ denotes an $SU(2)$ matrix whose phase describes the pions, $\nabla_0\Sigma = \partial_0\Sigma -\frac{i}{2}\mu_{I5}(\tau_3\Sigma+\Sigma\tau_3)$ is the covariant derivative, $\mu_{I5}$ denotes the axial isospin chemical potential and the trace is taken in flavor space. Note that in the presence of a thermal medium Lorentz invariance is broken resulting in two independent decay constants which are related through the pion velocity \cite{Son:2001ff},
\begin{align}
    \label{eq:relation f and f_t}
    u = \frac{f_{\pi}}{f_t}\,.
\end{align}

\section{Error analysis}
\label{app:autocorr}

\begin{figure}[tp]
	\includegraphics[scale=0.48]{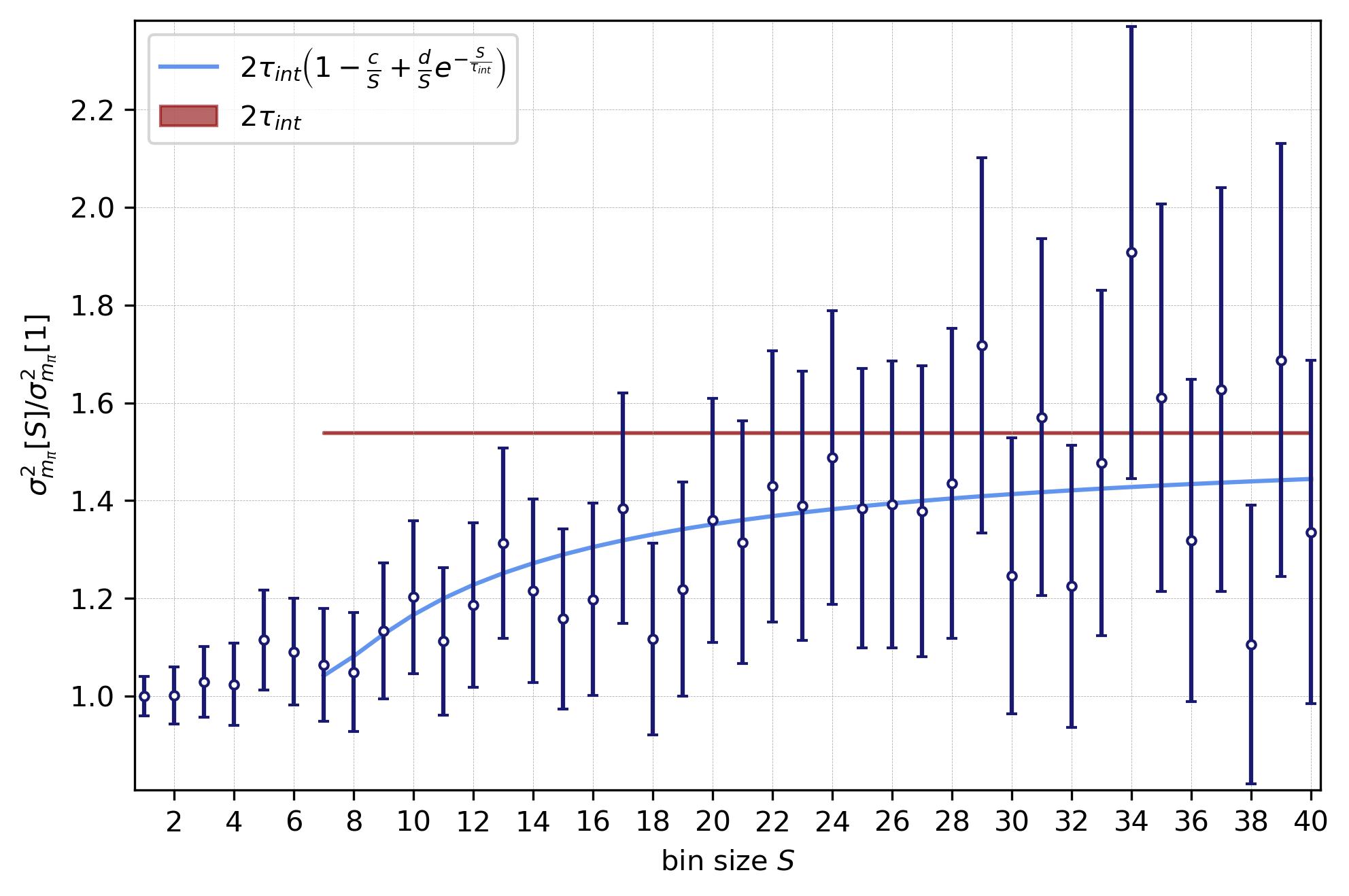}
	\includegraphics[scale=0.48]{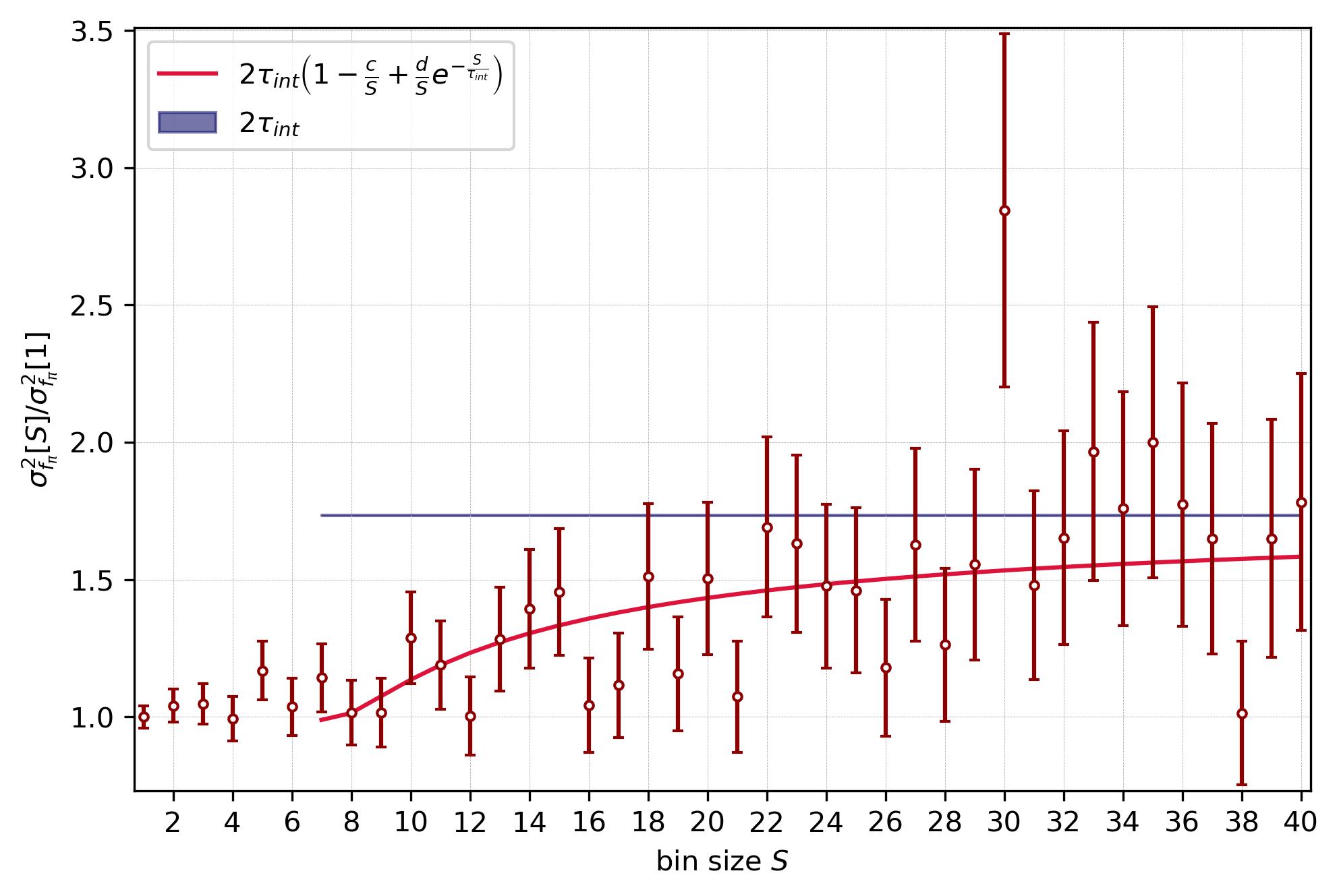}
\caption{
{\bfseries Left panel}: Normalized variance $\sigma^2_{m_{\pi}}[S]/\sigma^2_{m_{\pi}}[1]$ of the screening pion mass $m_{\pi}$ in dependence of the bin size $S$.
{\bfseries Right panel}: Normalized variance $\sigma^2_{f_{\pi}}[S]/\sigma^2_{f_{\pi}}[1]$ of the screening decay constant $f_{\pi}$ in dependence of the bin size $S$.
In both cases the fit function as well as its ``infinite bin size extrapolation'' for the integrated autocorrelation time $2\tau_{\text{int}}$ is shown.
}
\label{fig:autocorr}
\end{figure}

If $N_{\text{con.}} = 1200$ denotes the number of configurations, the mean value $\bar{O}$ of any lattice observable $O$ can be obtained via
\begin{align}
    \label{eq:mean_observable}
    \bar{O}=\frac{1}{N_{\text{con.}}}\sum_{i=1}^{N_{\text{con.}}}O_i\,.
\end{align}
All errors quoted in this work are purely statistical and estimated using \textit{jackknife} resampling \cite{Quenouille:1956:NBE}, where one first generates $N_{\text{con.}}$  jackknife replica  
\begin{align}
    \label{eq:jacknife_replica}
    O_i^J = \frac{1}{N_{\text{con.}}-1}\sum_{j\neq i}^{N_{\text{con.}}}O_j\,.
\end{align}
Employing this procedure the error on the mean of any lattice observable can be calculated as
\begin{align}
    \label{eq:jackknife_error}
    \sigma_{\bar{O}}=\sqrt{\frac{N_{\text{con.}}-1}{N_{\text{con.}}}\sum_i\left(O_i^J-\bar{O}\right)^2}\,,
\end{align}
where the additional factor $(N_{\text{con.}}-1)$ arises due to the fact that the jackknife replicas are not statistically independent.
Furthermore Eq.\,(\ref{eq:jackknife_error}) assumes uncorrelated jackknife replicas. However, since the configurations are obtained from Monte Carlo simulations the number of effectively independent jackknife replicas is \cite{Gattringer:2010zz}
\begin{align}
    N_{\text{ind.}} = \frac{N_{\text{con.}}}{2\tau_{O,\text{int}}}\,.
\end{align}
Following Ref.\,\cite{Bouma:2022teu}, we estimate the integrated autocorrelation time $2\tau_{O,\text{int}}$ as a fit parameter for the normalized variance using the fit ansatz
\begin{align}
    \label{eq:normalized_variance_fit}
   \frac{\sigma^2_O[S]}{\sigma^2_O[1]} = 2\tau_{O,\text{int}}\left(1-\frac{c}{S}+\frac{d}{S}e^{-\frac{S}{\tau_{O,\text{int}}}}\right)\,,\ \ \ O\in \{m_{\pi}, f_{\pi}\}\,.
\end{align}
It corresponds to an asymptote for infinite bin size $S$ (see Fig.\,\ref{fig:autocorr}).
Therefore - to avoid an underestimation of the error - we quote
\begin{align}
    \label{eq:error_including_autocorr}
    \hat{\sigma}_{\bar{O}} = \sqrt{2\tau_{O,\text{int}}}\,\sigma_{\bar{O}}\,,\ \ \ O\in \{m_{\pi}, f_{\pi}\}
\end{align}
as our final result.
For the remaining observables discussed in this work, autocorrelation effects were taken into account using binning, i.e. averaging the data samples over a bin size $N_{\text{bin}} = 20$,
\begin{align}
    \label{eq:binning}
    O_i^B = \frac{1}{N_{\text{bin}}}\sum_{j=N_{\text{bin}}(i-1)+1}^{iN_{\text{bin}}} O_j\,, \ \ \ i\in\left\{1,\dots,\frac{N_{\text{con.}}}{N_{\text{bin}}}=60\right\}\,,
\end{align}
before building the jackknife replicas $O_i^J$.
Therefore, in this case, one has to replace $O_j \rightarrow O_j^B$ and $N_{\text{con.}} \rightarrow N_{\text{con.}}/N_{\text{bin}}$ in  Eq.\,(\ref{eq:jacknife_replica}).

\section{Numerical values for temporal correlators}

In this appendix we list the means and errors of the (anti)symmetrized temporal correlators used in this work.

\begin{table}[tb]
\caption{Temporal (anti)symmetrized correlators projected to zero momentum. All errors quoted are purely statistical. Since only the ratios are needed the correlators are NOT renormalized.} 
\begin{tabular}{c c c c c}
\vspace{-0.3cm}
\\
\hline
\hline
\vspace{0.1cm}
$x_0/a$ & $\quad\quad G_{(V-A)_{\text{imp}}}(x_0,T)$ & $\quad\quad G_{(V-A)_{\text{rec}}}(x_0,T)$ & $\quad\quad G_{(PA_0)_{\text{imp}}}(x_0,T)$ & $\quad\quad G_{(PA_0)_{\text{rec}}}(x_0,T)$\\ \hline \vspace{0.1cm}
1 & $1.754(1)\times 10^{-2}$ & $1.76(3)\times 10^{-2}$   & $2.323(5)\times 10^{-2}$ & $2.35(4)\times 10^{-2}$ \\
2 & $1.392(3)\times 10^{-3}$ & $1.43(3)\times 10^{-3}$  & $4.63(2)\times 10^{-3}$ & $4.92(9)\times 10^{-3}$ \\
3 & $2.78(2)\times 10^{-4}$ & $3.09(6)\times 10^{-4}$  & $1.58(2)\times 10^{-3}$ & $1.85(4)\times 10^{-3}$ \\
4 & $9.7(2)\times 10^{-5}$ & $1.25(2)\times 10^{-4}$  & $1.06(2)\times 10^{-3}$ & $1.31(3)\times 10^{-3}$\\
5 & $6.1(2)\times 10^{-5}$ &  $8.6(2)\times 10^{-5}$ & $8.7(1)\times 10^{-4}$ & $1.09(2)\times 10^{-3}$ \\
6 & $4.9(2)\times 10^{-5}$ & $7.1(2)\times 10^{-5}$  & $7.3(1)\times 10^{-4}$ & $9.2(2)\times 10^{-4}$ \\
7 & $4.2(2)\times 10^{-5}$ & $6.2(1)\times 10^{-5}$  & $6.1(1)\times 10^{-4}$ & $7.7(2)\times 10^{-4}$ \\
8 & $3.8(2)\times 10^{-5}$& $5.5(1)\times 10^{-5}$  & $4.9(1)\times 10^{-4}$ & $6.1(2)\times 10^{-4}$ \\
9 & $3.4(2)\times 10^{-5}$ & $5.0(1)\times 10^{-5}$  & $3.7(1)\times 10^{-4}$ & $4.6(1)\times 10^{-4}$ \\
10 & $3.1(2)\times 10^{-5}$ & $4.6(1)\times 10^{-5}$  & $2.5(1)\times 10^{-4}$ & $3.0(1)\times 10^{-4}$ \\
11 & $2.9(2)\times 10^{-5}$ & $4.3(1)\times 10^{-5}$  & $1.2(1)\times 10^{-4}$ & $1.51(5)\times 10^{-4}$\\
12 & $2.8(2)\times 10^{-5}$ & $4.2(1)\times 10^{-5}$  & $0$ & $0$\\
\hline
\hline
\end{tabular}
\label{tab:temporal_correlators}
\end{table}


\bibliography{pion_quasiparticle.bib}

\begin{thebibliography}{51}%
\makeatletter
\providecommand \@ifxundefined [1]{%
 \@ifx{#1\undefined}
}%
\providecommand \@ifnum [1]{%
 \ifnum #1\expandafter \@firstoftwo
 \else \expandafter \@secondoftwo
 \fi
}%
\providecommand \@ifx [1]{%
 \ifx #1\expandafter \@firstoftwo
 \else \expandafter \@secondoftwo
 \fi
}%
\providecommand \natexlab [1]{#1}%
\providecommand \enquote  [1]{``#1''}%
\providecommand \bibnamefont  [1]{#1}%
\providecommand \bibfnamefont [1]{#1}%
\providecommand \citenamefont [1]{#1}%
\providecommand \href@noop [0]{\@secondoftwo}%
\providecommand \href [0]{\begingroup \@sanitize@url \@href}%
\providecommand \@href[1]{\@@startlink{#1}\@@href}%
\providecommand \@@href[1]{\endgroup#1\@@endlink}%
\providecommand \@sanitize@url [0]{\catcode `\\12\catcode `\$12\catcode
  `\&12\catcode `\#12\catcode `\^12\catcode `\_12\catcode `\%12\relax}%
\providecommand \@@startlink[1]{}%
\providecommand \@@endlink[0]{}%
\providecommand \url  [0]{\begingroup\@sanitize@url \@url }%
\providecommand \@url [1]{\endgroup\@href {#1}{\urlprefix }}%
\providecommand \urlprefix  [0]{URL }%
\providecommand \Eprint [0]{\href }%
\providecommand \doibase [0]{http://dx.doi.org/}%
\providecommand \selectlanguage [0]{\@gobble}%
\providecommand \bibinfo  [0]{\@secondoftwo}%
\providecommand \bibfield  [0]{\@secondoftwo}%
\providecommand \translation [1]{[#1]}%
\providecommand \BibitemOpen [0]{}%
\providecommand \bibitemStop [0]{}%
\providecommand \bibitemNoStop [0]{.\EOS\space}%
\providecommand \EOS [0]{\spacefactor3000\relax}%
\providecommand \BibitemShut  [1]{\csname bibitem#1\endcsname}%
\let\auto@bib@innerbib\@empty
\bibitem [{\citenamefont {Shuryak}(1990)}]{Shuryak:1990ie}%
  \BibitemOpen
  \bibfield  {author} {\bibinfo {author} {\bibfnamefont {E.~V.}\ \bibnamefont
  {Shuryak}},\ }\href {\doibase 10.1103/PhysRevD.42.1764} {\bibfield  {journal}
  {\bibinfo  {journal} {Phys. Rev. D}\ }\textbf {\bibinfo {volume} {42}},\
  \bibinfo {pages} {1764} (\bibinfo {year} {1990})}\BibitemShut {NoStop}%
\bibitem [{\citenamefont {Goity}\ and\ \citenamefont
  {Leutwyler}(1989)}]{Goity:1989gs}%
  \BibitemOpen
  \bibfield  {author} {\bibinfo {author} {\bibfnamefont {J.}~\bibnamefont
  {Goity}}\ and\ \bibinfo {author} {\bibfnamefont {H.}~\bibnamefont
  {Leutwyler}},\ }\href {\doibase 10.1016/0370-2693(89)90985-4} {\bibfield
  {journal} {\bibinfo  {journal} {Phys.Lett.}\ }\textbf {\bibinfo {volume}
  {B228}},\ \bibinfo {pages} {517} (\bibinfo {year} {1989})}\BibitemShut
  {NoStop}%
\bibitem [{\citenamefont {Gasser}\ and\ \citenamefont
  {Leutwyler}(1987)}]{Gasser:1987ah}%
  \BibitemOpen
  \bibfield  {author} {\bibinfo {author} {\bibfnamefont {J.}~\bibnamefont
  {Gasser}}\ and\ \bibinfo {author} {\bibfnamefont {H.}~\bibnamefont
  {Leutwyler}},\ }\href {\doibase 10.1016/0370-2693(87)91652-2} {\bibfield
  {journal} {\bibinfo  {journal} {Phys. Lett. B}\ }\textbf {\bibinfo {volume}
  {188}},\ \bibinfo {pages} {477} (\bibinfo {year} {1987})}\BibitemShut
  {NoStop}%
\bibitem [{\citenamefont {Gerber}\ and\ \citenamefont
  {Leutwyler}(1989)}]{Gerber:1988tt}%
  \BibitemOpen
  \bibfield  {author} {\bibinfo {author} {\bibfnamefont {P.}~\bibnamefont
  {Gerber}}\ and\ \bibinfo {author} {\bibfnamefont {H.}~\bibnamefont
  {Leutwyler}},\ }\href {\doibase 10.1016/0550-3213(89)90349-0} {\bibfield
  {journal} {\bibinfo  {journal} {Nucl. Phys. B}\ }\textbf {\bibinfo {volume}
  {321}},\ \bibinfo {pages} {387} (\bibinfo {year} {1989})}\BibitemShut
  {NoStop}%
\bibitem [{\citenamefont {Schenk}(1993)}]{Schenk:1993ru}%
  \BibitemOpen
  \bibfield  {author} {\bibinfo {author} {\bibfnamefont {A.}~\bibnamefont
  {Schenk}},\ }\href {\doibase 10.1103/PhysRevD.47.5138} {\bibfield  {journal}
  {\bibinfo  {journal} {Phys.Rev.}\ }\textbf {\bibinfo {volume} {D47}},\
  \bibinfo {pages} {5138} (\bibinfo {year} {1993})}\BibitemShut {NoStop}%
\bibitem [{\citenamefont {Toublan}(1997)}]{Toublan:1997rr}%
  \BibitemOpen
  \bibfield  {author} {\bibinfo {author} {\bibfnamefont {D.}~\bibnamefont
  {Toublan}},\ }\href {\doibase 10.1103/PhysRevD.56.5629} {\bibfield  {journal}
  {\bibinfo  {journal} {Phys.Rev.}\ }\textbf {\bibinfo {volume} {D56}},\
  \bibinfo {pages} {5629} (\bibinfo {year} {1997})},\ \Eprint
  {http://arxiv.org/abs/hep-ph/9706273} {arXiv:hep-ph/9706273 [hep-ph]}
  \BibitemShut {NoStop}%
\bibitem [{\citenamefont {Pisarski}\ and\ \citenamefont
  {Tytgat}(1996)}]{Pisarski:1996mt}%
  \BibitemOpen
  \bibfield  {author} {\bibinfo {author} {\bibfnamefont {R.~D.}\ \bibnamefont
  {Pisarski}}\ and\ \bibinfo {author} {\bibfnamefont {M.}~\bibnamefont
  {Tytgat}},\ }\href {\doibase 10.1103/PhysRevD.54.R2989} {\bibfield  {journal}
  {\bibinfo  {journal} {Phys.Rev.}\ }\textbf {\bibinfo {volume} {D54}},\
  \bibinfo {pages} {2989} (\bibinfo {year} {1996})},\ \Eprint
  {http://arxiv.org/abs/hep-ph/9604404} {arXiv:hep-ph/9604404 [hep-ph]}
  \BibitemShut {NoStop}%
\bibitem [{\citenamefont {Son}\ and\ \citenamefont
  {Stephanov}(2002{\natexlab{a}})}]{Son:2001ff}%
  \BibitemOpen
  \bibfield  {author} {\bibinfo {author} {\bibfnamefont {D.~T.}\ \bibnamefont
  {Son}}\ and\ \bibinfo {author} {\bibfnamefont {M.~A.}\ \bibnamefont
  {Stephanov}},\ }\href {\doibase 10.1103/PhysRevLett.88.202302} {\bibfield
  {journal} {\bibinfo  {journal} {Phys. Rev. Lett.}\ }\textbf {\bibinfo
  {volume} {88}},\ \bibinfo {pages} {202302} (\bibinfo {year}
  {2002}{\natexlab{a}})},\ \Eprint {http://arxiv.org/abs/hep-ph/0111100}
  {arXiv:hep-ph/0111100} \BibitemShut {NoStop}%
\bibitem [{\citenamefont {Son}\ and\ \citenamefont
  {Stephanov}(2002{\natexlab{b}})}]{Son:2002ci}%
  \BibitemOpen
  \bibfield  {author} {\bibinfo {author} {\bibfnamefont {D.~T.}\ \bibnamefont
  {Son}}\ and\ \bibinfo {author} {\bibfnamefont {M.~A.}\ \bibnamefont
  {Stephanov}},\ }\href {\doibase 10.1103/PhysRevD.66.076011} {\bibfield
  {journal} {\bibinfo  {journal} {Phys. Rev. D}\ }\textbf {\bibinfo {volume}
  {66}},\ \bibinfo {pages} {076011} (\bibinfo {year} {2002}{\natexlab{b}})},\
  \Eprint {http://arxiv.org/abs/hep-ph/0204226} {arXiv:hep-ph/0204226}
  \BibitemShut {NoStop}%
\bibitem [{\citenamefont {Brandt}\ \emph {et~al.}(2014)\citenamefont {Brandt},
  \citenamefont {Francis}, \citenamefont {Meyer},\ and\ \citenamefont
  {Robaina}}]{Brandt:2014qqa}%
  \BibitemOpen
  \bibfield  {author} {\bibinfo {author} {\bibfnamefont {B.~B.}\ \bibnamefont
  {Brandt}}, \bibinfo {author} {\bibfnamefont {A.}~\bibnamefont {Francis}},
  \bibinfo {author} {\bibfnamefont {H.~B.}\ \bibnamefont {Meyer}}, \ and\
  \bibinfo {author} {\bibfnamefont {D.}~\bibnamefont {Robaina}},\ }\href
  {\doibase 10.1103/PhysRevD.90.054509} {\bibfield  {journal} {\bibinfo
  {journal} {Phys. Rev. D}\ }\textbf {\bibinfo {volume} {90}},\ \bibinfo
  {pages} {054509} (\bibinfo {year} {2014})},\ \Eprint
  {http://arxiv.org/abs/1406.5602} {arXiv:1406.5602 [hep-lat]} \BibitemShut
  {NoStop}%
\bibitem [{\citenamefont {Brandt}\ \emph {et~al.}(2015)\citenamefont {Brandt},
  \citenamefont {Francis}, \citenamefont {Meyer},\ and\ \citenamefont
  {Robaina}}]{Brandt:2015sxa}%
  \BibitemOpen
  \bibfield  {author} {\bibinfo {author} {\bibfnamefont {B.~B.}\ \bibnamefont
  {Brandt}}, \bibinfo {author} {\bibfnamefont {A.}~\bibnamefont {Francis}},
  \bibinfo {author} {\bibfnamefont {H.~B.}\ \bibnamefont {Meyer}}, \ and\
  \bibinfo {author} {\bibfnamefont {D.}~\bibnamefont {Robaina}},\ }\href
  {\doibase 10.1103/PhysRevD.92.094510} {\bibfield  {journal} {\bibinfo
  {journal} {Phys. Rev. D}\ }\textbf {\bibinfo {volume} {92}},\ \bibinfo
  {pages} {094510} (\bibinfo {year} {2015})},\ \Eprint
  {http://arxiv.org/abs/1506.05732} {arXiv:1506.05732 [hep-lat]} \BibitemShut
  {NoStop}%
\bibitem [{\citenamefont {L{\"u}scher}(1998)}]{Luscher:1998pe}%
  \BibitemOpen
  \bibfield  {author} {\bibinfo {author} {\bibfnamefont {M.}~\bibnamefont
  {L{\"u}scher}},\ }in\ \href@noop {} {\emph {\bibinfo {booktitle} {{Les
  Houches Summer School in Theoretical Physics, Session 68: Probing the
  Standard Model of Particle Interactions}}}}\ (\bibinfo {year} {1998})\ pp.\
  \bibinfo {pages} {229--280},\ \Eprint {http://arxiv.org/abs/hep-lat/9802029}
  {arXiv:hep-lat/9802029} \BibitemShut {NoStop}%
\bibitem [{\citenamefont {Meyer}(2011)}]{Meyer:2011gj}%
  \BibitemOpen
  \bibfield  {author} {\bibinfo {author} {\bibfnamefont {H.~B.}\ \bibnamefont
  {Meyer}},\ }\href {\doibase 10.1140/epja/i2011-11086-3} {\bibfield  {journal}
  {\bibinfo  {journal} {Eur. Phys. J. A}\ }\textbf {\bibinfo {volume} {47}},\
  \bibinfo {pages} {86} (\bibinfo {year} {2011})},\ \Eprint
  {http://arxiv.org/abs/1104.3708} {arXiv:1104.3708 [hep-lat]} \BibitemShut
  {NoStop}%
\bibitem [{\citenamefont {Bulava}\ \emph {et~al.}(2015)\citenamefont {Bulava},
  \citenamefont {Della~Morte}, \citenamefont {Heitger},\ and\ \citenamefont
  {Wittemeier}}]{Bulava:2015bxa}%
  \BibitemOpen
  \bibfield  {author} {\bibinfo {author} {\bibfnamefont {J.}~\bibnamefont
  {Bulava}}, \bibinfo {author} {\bibfnamefont {M.}~\bibnamefont {Della~Morte}},
  \bibinfo {author} {\bibfnamefont {J.}~\bibnamefont {Heitger}}, \ and\
  \bibinfo {author} {\bibfnamefont {C.}~\bibnamefont {Wittemeier}} (\bibinfo
  {collaboration} {ALPHA}),\ }\href {\doibase 10.1016/j.nuclphysb.2015.05.003}
  {\bibfield  {journal} {\bibinfo  {journal} {Nucl. Phys. B}\ }\textbf
  {\bibinfo {volume} {896}},\ \bibinfo {pages} {555} (\bibinfo {year}
  {2015})},\ \Eprint {http://arxiv.org/abs/1502.04999} {arXiv:1502.04999
  [hep-lat]} \BibitemShut {NoStop}%
\bibitem [{\citenamefont {Gerardin}\ \emph {et~al.}(2019)\citenamefont
  {Gerardin}, \citenamefont {Harris},\ and\ \citenamefont
  {Meyer}}]{Gerardin:2018kpy}%
  \BibitemOpen
  \bibfield  {author} {\bibinfo {author} {\bibfnamefont {A.}~\bibnamefont
  {Gerardin}}, \bibinfo {author} {\bibfnamefont {T.}~\bibnamefont {Harris}}, \
  and\ \bibinfo {author} {\bibfnamefont {H.~B.}\ \bibnamefont {Meyer}},\ }\href
  {\doibase 10.1103/PhysRevD.99.014519} {\bibfield  {journal} {\bibinfo
  {journal} {Phys. Rev. D}\ }\textbf {\bibinfo {volume} {99}},\ \bibinfo
  {pages} {014519} (\bibinfo {year} {2019})},\ \Eprint
  {http://arxiv.org/abs/1811.08209} {arXiv:1811.08209 [hep-lat]} \BibitemShut
  {NoStop}%
\bibitem [{\citenamefont {Campos}\ \emph {et~al.}(2018)\citenamefont {Campos},
  \citenamefont {Fritzsch}, \citenamefont {Pena}, \citenamefont {Preti},
  \citenamefont {Ramos},\ and\ \citenamefont {Vladikas}}]{Campos:2018ahf}%
  \BibitemOpen
  \bibfield  {author} {\bibinfo {author} {\bibfnamefont {I.}~\bibnamefont
  {Campos}}, \bibinfo {author} {\bibfnamefont {P.}~\bibnamefont {Fritzsch}},
  \bibinfo {author} {\bibfnamefont {C.}~\bibnamefont {Pena}}, \bibinfo {author}
  {\bibfnamefont {D.}~\bibnamefont {Preti}}, \bibinfo {author} {\bibfnamefont
  {A.}~\bibnamefont {Ramos}}, \ and\ \bibinfo {author} {\bibfnamefont
  {A.}~\bibnamefont {Vladikas}} (\bibinfo {collaboration} {ALPHA}),\ }\href
  {\doibase 10.1140/epjc/s10052-018-5870-5} {\bibfield  {journal} {\bibinfo
  {journal} {Eur. Phys. J. C}\ }\textbf {\bibinfo {volume} {78}},\ \bibinfo
  {pages} {387} (\bibinfo {year} {2018})},\ \Eprint
  {http://arxiv.org/abs/1802.05243} {arXiv:1802.05243 [hep-lat]} \BibitemShut
  {NoStop}%
\bibitem [{\citenamefont {Bulava}\ and\ \citenamefont
  {Schaefer}(2013)}]{Bulava:2013cta}%
  \BibitemOpen
  \bibfield  {author} {\bibinfo {author} {\bibfnamefont {J.}~\bibnamefont
  {Bulava}}\ and\ \bibinfo {author} {\bibfnamefont {S.}~\bibnamefont
  {Schaefer}},\ }\href {\doibase 10.1016/j.nuclphysb.2013.05.019} {\bibfield
  {journal} {\bibinfo  {journal} {Nucl. Phys. B}\ }\textbf {\bibinfo {volume}
  {874}},\ \bibinfo {pages} {188} (\bibinfo {year} {2013})},\ \Eprint
  {http://arxiv.org/abs/1304.7093} {arXiv:1304.7093 [hep-lat]} \BibitemShut
  {NoStop}%
\bibitem [{\citenamefont {Bruno}\ \emph {et~al.}(2015)\citenamefont {Bruno}
  \emph {et~al.}}]{Bruno:2014jqa}%
  \BibitemOpen
  \bibfield  {author} {\bibinfo {author} {\bibfnamefont {M.}~\bibnamefont
  {Bruno}} \emph {et~al.},\ }\href {\doibase 10.1007/JHEP02(2015)043}
  {\bibfield  {journal} {\bibinfo  {journal} {JHEP}\ }\textbf {\bibinfo
  {volume} {02}},\ \bibinfo {pages} {043} (\bibinfo {year} {2015})},\ \Eprint
  {http://arxiv.org/abs/1411.3982} {arXiv:1411.3982 [hep-lat]} \BibitemShut
  {NoStop}%
\bibitem [{\citenamefont {Mohler}\ \emph {et~al.}(2018)\citenamefont {Mohler},
  \citenamefont {Schaefer},\ and\ \citenamefont {Simeth}}]{Mohler:2017wnb}%
  \BibitemOpen
  \bibfield  {author} {\bibinfo {author} {\bibfnamefont {D.}~\bibnamefont
  {Mohler}}, \bibinfo {author} {\bibfnamefont {S.}~\bibnamefont {Schaefer}}, \
  and\ \bibinfo {author} {\bibfnamefont {J.}~\bibnamefont {Simeth}},\ }\href
  {\doibase 10.1051/epjconf/201817502010} {\bibfield  {journal} {\bibinfo
  {journal} {EPJ Web Conf.}\ }\textbf {\bibinfo {volume} {175}},\ \bibinfo
  {pages} {02010} (\bibinfo {year} {2018})},\ \Eprint
  {http://arxiv.org/abs/1712.04884} {arXiv:1712.04884 [hep-lat]} \BibitemShut
  {NoStop}%
\bibitem [{\citenamefont {Bruno}\ \emph {et~al.}(2017)\citenamefont {Bruno},
  \citenamefont {Korzec},\ and\ \citenamefont {Schaefer}}]{Bruno:2016plf}%
  \BibitemOpen
  \bibfield  {author} {\bibinfo {author} {\bibfnamefont {M.}~\bibnamefont
  {Bruno}}, \bibinfo {author} {\bibfnamefont {T.}~\bibnamefont {Korzec}}, \
  and\ \bibinfo {author} {\bibfnamefont {S.}~\bibnamefont {Schaefer}},\ }\href
  {\doibase 10.1103/PhysRevD.95.074504} {\bibfield  {journal} {\bibinfo
  {journal} {Phys. Rev.}\ }\textbf {\bibinfo {volume} {D95}},\ \bibinfo {pages}
  {074504} (\bibinfo {year} {2017})},\ \Eprint
  {http://arxiv.org/abs/1608.08900} {arXiv:1608.08900 [hep-lat]} \BibitemShut
  {NoStop}%
\bibitem [{\citenamefont {Aoki}\ \emph {et~al.}(2006)\citenamefont {Aoki},
  \citenamefont {Fodor}, \citenamefont {Katz},\ and\ \citenamefont
  {Szabo}}]{Aoki:2006br}%
  \BibitemOpen
  \bibfield  {author} {\bibinfo {author} {\bibfnamefont {Y.}~\bibnamefont
  {Aoki}}, \bibinfo {author} {\bibfnamefont {Z.}~\bibnamefont {Fodor}},
  \bibinfo {author} {\bibfnamefont {S.~D.}\ \bibnamefont {Katz}}, \ and\
  \bibinfo {author} {\bibfnamefont {K.~K.}\ \bibnamefont {Szabo}},\ }\href
  {\doibase 10.1016/j.physletb.2006.10.021} {\bibfield  {journal} {\bibinfo
  {journal} {Phys. Lett.}\ }\textbf {\bibinfo {volume} {B643}},\ \bibinfo
  {pages} {46} (\bibinfo {year} {2006})},\ \Eprint
  {http://arxiv.org/abs/hep-lat/0609068} {arXiv:hep-lat/0609068} \BibitemShut
  {NoStop}%
\bibitem [{\citenamefont {Aoki}\ \emph {et~al.}(2009)\citenamefont {Aoki} \emph
  {et~al.}}]{Aoki:2009sc}%
  \BibitemOpen
  \bibfield  {author} {\bibinfo {author} {\bibfnamefont {Y.}~\bibnamefont
  {Aoki}} \emph {et~al.},\ }\href {\doibase 10.1088/1126-6708/2009/06/088}
  {\bibfield  {journal} {\bibinfo  {journal} {JHEP}\ }\textbf {\bibinfo
  {volume} {06}},\ \bibinfo {pages} {088} (\bibinfo {year} {2009})},\ \Eprint
  {http://arxiv.org/abs/0903.4155} {arXiv:0903.4155 [hep-lat]} \BibitemShut
  {NoStop}%
\bibitem [{\citenamefont {Borsanyi}\ \emph {et~al.}(2014)\citenamefont
  {Borsanyi}, \citenamefont {Fodor}, \citenamefont {Hoelbling}, \citenamefont
  {Katz}, \citenamefont {Krieg},\ and\ \citenamefont
  {Szabo}}]{Borsanyi:2013bia}%
  \BibitemOpen
  \bibfield  {author} {\bibinfo {author} {\bibfnamefont {S.}~\bibnamefont
  {Borsanyi}}, \bibinfo {author} {\bibfnamefont {Z.}~\bibnamefont {Fodor}},
  \bibinfo {author} {\bibfnamefont {C.}~\bibnamefont {Hoelbling}}, \bibinfo
  {author} {\bibfnamefont {S.~D.}\ \bibnamefont {Katz}}, \bibinfo {author}
  {\bibfnamefont {S.}~\bibnamefont {Krieg}}, \ and\ \bibinfo {author}
  {\bibfnamefont {K.~K.}\ \bibnamefont {Szabo}},\ }\href {\doibase
  10.1016/j.physletb.2014.01.007} {\bibfield  {journal} {\bibinfo  {journal}
  {Phys. Lett. B}\ }\textbf {\bibinfo {volume} {730}},\ \bibinfo {pages} {99}
  (\bibinfo {year} {2014})},\ \Eprint {http://arxiv.org/abs/1309.5258}
  {arXiv:1309.5258 [hep-lat]} \BibitemShut {NoStop}%
\bibitem [{\citenamefont {C\`e}\ \emph
  {et~al.}(2022{\natexlab{a}})\citenamefont {C\`e} \emph
  {et~al.}}]{Ce:2022kxy}%
  \BibitemOpen
  \bibfield  {author} {\bibinfo {author} {\bibfnamefont {M.}~\bibnamefont
  {C\`e}} \emph {et~al.},\ }\href@noop {} {\  (\bibinfo {year}
  {2022}{\natexlab{a}})},\ \Eprint {http://arxiv.org/abs/2206.06582}
  {arXiv:2206.06582 [hep-lat]} \BibitemShut {NoStop}%
\bibitem [{\citenamefont {L{\"u}scher}\ and\ \citenamefont
  {Schaefer}(2013)}]{Luscher:2012av}%
  \BibitemOpen
  \bibfield  {author} {\bibinfo {author} {\bibfnamefont {M.}~\bibnamefont
  {L{\"u}scher}}\ and\ \bibinfo {author} {\bibfnamefont {S.}~\bibnamefont
  {Schaefer}},\ }\href {\doibase 10.1016/j.cpc.2012.10.003} {\bibfield
  {journal} {\bibinfo  {journal} {Comput. Phys. Commun.}\ }\textbf {\bibinfo
  {volume} {184}},\ \bibinfo {pages} {519} (\bibinfo {year} {2013})},\ \Eprint
  {http://arxiv.org/abs/1206.2809} {arXiv:1206.2809 [hep-lat]} \BibitemShut
  {NoStop}%
\bibitem [{\citenamefont {Press}\ \emph {et~al.}(2007)\citenamefont {Press},
  \citenamefont {Teukolsky}, \citenamefont {Vetterling},\ and\ \citenamefont
  {Flannery}}]{press2007numerical}%
  \BibitemOpen
  \bibfield  {author} {\bibinfo {author} {\bibfnamefont {W.}~\bibnamefont
  {Press}}, \bibinfo {author} {\bibfnamefont {S.}~\bibnamefont {Teukolsky}},
  \bibinfo {author} {\bibfnamefont {W.}~\bibnamefont {Vetterling}}, \ and\
  \bibinfo {author} {\bibfnamefont {B.}~\bibnamefont {Flannery}},\ }\href
  {http://nr.com/} {\emph {\bibinfo {title} {Numerical Recipes: The Art of
  Scientific Computing}}},\ \bibinfo {edition} {3rd}\ ed.\ (\bibinfo
  {publisher} {Cambridge University Press},\ \bibinfo {year}
  {2007})\BibitemShut {NoStop}%
\bibitem [{\citenamefont {C\`e}\ \emph
  {et~al.}(2022{\natexlab{b}})\citenamefont {C\`e}, \citenamefont {G\'erardin},
  \citenamefont {von Hippel}, \citenamefont {Meyer}, \citenamefont {Miura},
  \citenamefont {Ottnad}, \citenamefont {Risch}, \citenamefont {San~Jos\'e},
  \citenamefont {Wilhelm},\ and\ \citenamefont {Wittig}}]{Ce:2022eix}%
  \BibitemOpen
  \bibfield  {author} {\bibinfo {author} {\bibfnamefont {M.}~\bibnamefont
  {C\`e}}, \bibinfo {author} {\bibfnamefont {A.}~\bibnamefont {G\'erardin}},
  \bibinfo {author} {\bibfnamefont {G.}~\bibnamefont {von Hippel}}, \bibinfo
  {author} {\bibfnamefont {H.~B.}\ \bibnamefont {Meyer}}, \bibinfo {author}
  {\bibfnamefont {K.}~\bibnamefont {Miura}}, \bibinfo {author} {\bibfnamefont
  {K.}~\bibnamefont {Ottnad}}, \bibinfo {author} {\bibfnamefont
  {A.}~\bibnamefont {Risch}}, \bibinfo {author} {\bibfnamefont
  {T.}~\bibnamefont {San~Jos\'e}}, \bibinfo {author} {\bibfnamefont
  {J.}~\bibnamefont {Wilhelm}}, \ and\ \bibinfo {author} {\bibfnamefont
  {H.}~\bibnamefont {Wittig}},\ }\href {\doibase 10.1007/JHEP08(2022)220}
  {\bibfield  {journal} {\bibinfo  {journal} {JHEP}\ }\textbf {\bibinfo
  {volume} {08}},\ \bibinfo {pages} {220} (\bibinfo {year}
  {2022}{\natexlab{b}})},\ \Eprint {http://arxiv.org/abs/2203.08676}
  {arXiv:2203.08676 [hep-lat]} \BibitemShut {NoStop}%
\bibitem [{\citenamefont {Backus}\ and\ \citenamefont
  {Gilbert}(1968)}]{Backus:1968}%
  \BibitemOpen
  \bibfield  {author} {\bibinfo {author} {\bibfnamefont {G.}~\bibnamefont
  {Backus}}\ and\ \bibinfo {author} {\bibfnamefont {F.}~\bibnamefont
  {Gilbert}},\ }\href {\doibase 10.1111/j.1365-246X.1968.tb00216.x} {\bibfield
  {journal} {\bibinfo  {journal} {Geophysical Journal International}\ }\textbf
  {\bibinfo {volume} {16}},\ \bibinfo {pages} {169} (\bibinfo {year} {1968})},\
  \Eprint
  {http://arxiv.org/abs/https://academic.oup.com/gji/article-pdf/16/2/169/5891044/16-2-169.pdf}
  {https://academic.oup.com/gji/article-pdf/16/2/169/5891044/16-2-169.pdf}
  \BibitemShut {NoStop}%
\bibitem [{\citenamefont {Bazavov}\ \emph {et~al.}(2019)\citenamefont {Bazavov}
  \emph {et~al.}}]{Bazavov:2019www}%
  \BibitemOpen
  \bibfield  {author} {\bibinfo {author} {\bibfnamefont {A.}~\bibnamefont
  {Bazavov}} \emph {et~al.},\ }\href {\doibase 10.1103/PhysRevD.100.094510}
  {\bibfield  {journal} {\bibinfo  {journal} {Phys. Rev. D}\ }\textbf {\bibinfo
  {volume} {100}},\ \bibinfo {pages} {094510} (\bibinfo {year} {2019})},\
  \Eprint {http://arxiv.org/abs/1908.09552} {arXiv:1908.09552 [hep-lat]}
  \BibitemShut {NoStop}%
\bibitem [{\citenamefont {Goderidze}\ \emph {et~al.}(2022)\citenamefont
  {Goderidze}, \citenamefont {Friesen},\ and\ \citenamefont
  {Kalinovsky}}]{Goderidze:2022vlm}%
  \BibitemOpen
  \bibfield  {author} {\bibinfo {author} {\bibfnamefont {D.}~\bibnamefont
  {Goderidze}}, \bibinfo {author} {\bibfnamefont {A.}~\bibnamefont {Friesen}},
  \ and\ \bibinfo {author} {\bibfnamefont {Y.}~\bibnamefont {Kalinovsky}},\
  }\href {\doibase 10.1142/S0217751X22501354} {\bibfield  {journal} {\bibinfo
  {journal} {Int. J. Mod. Phys. A}\ }\textbf {\bibinfo {volume} {37}},\
  \bibinfo {pages} {2250135} (\bibinfo {year} {2022})},\ \Eprint
  {http://arxiv.org/abs/2205.11436} {arXiv:2205.11436 [hep-ph]} \BibitemShut
  {NoStop}%
\bibitem [{\citenamefont {Borsanyi}\ \emph {et~al.}(2012)\citenamefont
  {Borsanyi}, \citenamefont {Fodor}, \citenamefont {Katz}, \citenamefont
  {Krieg}, \citenamefont {Ratti},\ and\ \citenamefont
  {Szabo}}]{Borsanyi:2011sw}%
  \BibitemOpen
  \bibfield  {author} {\bibinfo {author} {\bibfnamefont {S.}~\bibnamefont
  {Borsanyi}}, \bibinfo {author} {\bibfnamefont {Z.}~\bibnamefont {Fodor}},
  \bibinfo {author} {\bibfnamefont {S.~D.}\ \bibnamefont {Katz}}, \bibinfo
  {author} {\bibfnamefont {S.}~\bibnamefont {Krieg}}, \bibinfo {author}
  {\bibfnamefont {C.}~\bibnamefont {Ratti}}, \ and\ \bibinfo {author}
  {\bibfnamefont {K.}~\bibnamefont {Szabo}},\ }\href {\doibase
  10.1007/JHEP01(2012)138} {\bibfield  {journal} {\bibinfo  {journal} {JHEP}\
  }\textbf {\bibinfo {volume} {01}},\ \bibinfo {pages} {138} (\bibinfo {year}
  {2012})},\ \Eprint {http://arxiv.org/abs/1112.4416} {arXiv:1112.4416
  [hep-lat]} \BibitemShut {NoStop}%
\bibitem [{\citenamefont {Hagedorn}(1985)}]{Hagedorn:1984hz}%
  \BibitemOpen
  \bibfield  {author} {\bibinfo {author} {\bibfnamefont {R.}~\bibnamefont
  {Hagedorn}},\ }\href {\doibase 10.1007/978-3-319-17545-4_25} {\bibfield
  {journal} {\bibinfo  {journal} {Lect. Notes Phys.}\ }\textbf {\bibinfo
  {volume} {221}},\ \bibinfo {pages} {53} (\bibinfo {year} {1985})}\BibitemShut
  {NoStop}%
\bibitem [{\citenamefont {Karsch}\ \emph {et~al.}(2003)\citenamefont {Karsch},
  \citenamefont {Redlich},\ and\ \citenamefont {Tawfik}}]{Karsch:2003vd}%
  \BibitemOpen
  \bibfield  {author} {\bibinfo {author} {\bibfnamefont {F.}~\bibnamefont
  {Karsch}}, \bibinfo {author} {\bibfnamefont {K.}~\bibnamefont {Redlich}}, \
  and\ \bibinfo {author} {\bibfnamefont {A.}~\bibnamefont {Tawfik}},\ }\href
  {\doibase 10.1140/epjc/s2003-01228-y} {\bibfield  {journal} {\bibinfo
  {journal} {Eur. Phys. J. C}\ }\textbf {\bibinfo {volume} {29}},\ \bibinfo
  {pages} {549} (\bibinfo {year} {2003})},\ \Eprint
  {http://arxiv.org/abs/hep-ph/0303108} {arXiv:hep-ph/0303108} \BibitemShut
  {NoStop}%
\bibitem [{\citenamefont {Brandt}\ \emph {et~al.}(2016)\citenamefont {Brandt},
  \citenamefont {Francis}, \citenamefont {J\"ager},\ and\ \citenamefont
  {Meyer}}]{Brandt:2015aqk}%
  \BibitemOpen
  \bibfield  {author} {\bibinfo {author} {\bibfnamefont {B.~B.}\ \bibnamefont
  {Brandt}}, \bibinfo {author} {\bibfnamefont {A.}~\bibnamefont {Francis}},
  \bibinfo {author} {\bibfnamefont {B.}~\bibnamefont {J\"ager}}, \ and\
  \bibinfo {author} {\bibfnamefont {H.~B.}\ \bibnamefont {Meyer}},\ }\href
  {\doibase 10.1103/PhysRevD.93.054510} {\bibfield  {journal} {\bibinfo
  {journal} {Phys. Rev. D}\ }\textbf {\bibinfo {volume} {93}},\ \bibinfo
  {pages} {054510} (\bibinfo {year} {2016})},\ \Eprint
  {http://arxiv.org/abs/1512.07249} {arXiv:1512.07249 [hep-lat]} \BibitemShut
  {NoStop}%
\bibitem [{\citenamefont {Davier}\ \emph {et~al.}(2006)\citenamefont {Davier},
  \citenamefont {Hocker},\ and\ \citenamefont {Zhang}}]{Davier:2005xq}%
  \BibitemOpen
  \bibfield  {author} {\bibinfo {author} {\bibfnamefont {M.}~\bibnamefont
  {Davier}}, \bibinfo {author} {\bibfnamefont {A.}~\bibnamefont {Hocker}}, \
  and\ \bibinfo {author} {\bibfnamefont {Z.}~\bibnamefont {Zhang}},\ }\href
  {\doibase 10.1103/RevModPhys.78.1043} {\bibfield  {journal} {\bibinfo
  {journal} {Rev.Mod.Phys.}\ }\textbf {\bibinfo {volume} {78}},\ \bibinfo
  {pages} {1043} (\bibinfo {year} {2006})},\ \Eprint
  {http://arxiv.org/abs/hep-ph/0507078} {arXiv:hep-ph/0507078 [hep-ph]}
  \BibitemShut {NoStop}%
\bibitem [{\citenamefont {Rapp}\ and\ \citenamefont
  {Wambach}(2000)}]{Rapp:1999ej}%
  \BibitemOpen
  \bibfield  {author} {\bibinfo {author} {\bibfnamefont {R.}~\bibnamefont
  {Rapp}}\ and\ \bibinfo {author} {\bibfnamefont {J.}~\bibnamefont {Wambach}},\
  }\href {\doibase 10.1007/0-306-47101-9_1} {\bibfield  {journal} {\bibinfo
  {journal} {Adv. Nucl. Phys.}\ }\textbf {\bibinfo {volume} {25}},\ \bibinfo
  {pages} {1} (\bibinfo {year} {2000})},\ \Eprint
  {http://arxiv.org/abs/hep-ph/9909229} {arXiv:hep-ph/9909229} \BibitemShut
  {NoStop}%
\bibitem [{\citenamefont {Kapusta}\ and\ \citenamefont
  {Shuryak}(1994)}]{Kapusta:1993hq}%
  \BibitemOpen
  \bibfield  {author} {\bibinfo {author} {\bibfnamefont {J.~I.}\ \bibnamefont
  {Kapusta}}\ and\ \bibinfo {author} {\bibfnamefont {E.~V.}\ \bibnamefont
  {Shuryak}},\ }\href {\doibase 10.1103/PhysRevD.49.4694} {\bibfield  {journal}
  {\bibinfo  {journal} {Phys. Rev.}\ }\textbf {\bibinfo {volume} {D49}},\
  \bibinfo {pages} {4694} (\bibinfo {year} {1994})},\ \Eprint
  {http://arxiv.org/abs/hep-ph/9312245} {arXiv:hep-ph/9312245} \BibitemShut
  {NoStop}%
\bibitem [{\citenamefont {Hohler}\ and\ \citenamefont
  {Rapp}(2014)}]{Hohler:2013eba}%
  \BibitemOpen
  \bibfield  {author} {\bibinfo {author} {\bibfnamefont {P.~M.}\ \bibnamefont
  {Hohler}}\ and\ \bibinfo {author} {\bibfnamefont {R.}~\bibnamefont {Rapp}},\
  }\href {\doibase 10.1016/j.physletb.2014.02.021} {\bibfield  {journal}
  {\bibinfo  {journal} {Phys.Lett.}\ }\textbf {\bibinfo {volume} {B731}},\
  \bibinfo {pages} {103} (\bibinfo {year} {2014})},\ \Eprint
  {http://arxiv.org/abs/1311.2921} {arXiv:1311.2921 [hep-ph]} \BibitemShut
  {NoStop}%
\bibitem [{\citenamefont {Dey}\ \emph {et~al.}(1990)\citenamefont {Dey},
  \citenamefont {Eletsky},\ and\ \citenamefont {Ioffe}}]{Dey:1990ba}%
  \BibitemOpen
  \bibfield  {author} {\bibinfo {author} {\bibfnamefont {M.}~\bibnamefont
  {Dey}}, \bibinfo {author} {\bibfnamefont {V.~L.}\ \bibnamefont {Eletsky}}, \
  and\ \bibinfo {author} {\bibfnamefont {B.~L.}\ \bibnamefont {Ioffe}},\ }\href
  {\doibase 10.1016/0370-2693(90)90495-R} {\bibfield  {journal} {\bibinfo
  {journal} {Phys. Lett. B}\ }\textbf {\bibinfo {volume} {252}},\ \bibinfo
  {pages} {620} (\bibinfo {year} {1990})}\BibitemShut {NoStop}%
\bibitem [{\citenamefont {Eletsky}\ and\ \citenamefont
  {Ioffe}(1993)}]{Eletsky:1992ay}%
  \BibitemOpen
  \bibfield  {author} {\bibinfo {author} {\bibfnamefont {V.~L.}\ \bibnamefont
  {Eletsky}}\ and\ \bibinfo {author} {\bibfnamefont {B.~L.}\ \bibnamefont
  {Ioffe}},\ }\href {\doibase 10.1103/PhysRevD.47.3083} {\bibfield  {journal}
  {\bibinfo  {journal} {Phys. Rev. D}\ }\textbf {\bibinfo {volume} {47}},\
  \bibinfo {pages} {3083} (\bibinfo {year} {1993})},\ \Eprint
  {http://arxiv.org/abs/hep-ph/9302298} {arXiv:hep-ph/9302298} \BibitemShut
  {NoStop}%
\bibitem [{\citenamefont {Eletsky}\ and\ \citenamefont
  {Ioffe}(1995)}]{Eletsky:1994rp}%
  \BibitemOpen
  \bibfield  {author} {\bibinfo {author} {\bibfnamefont {V.~L.}\ \bibnamefont
  {Eletsky}}\ and\ \bibinfo {author} {\bibfnamefont {B.~L.}\ \bibnamefont
  {Ioffe}},\ }\href {\doibase 10.1103/PhysRevD.51.2371} {\bibfield  {journal}
  {\bibinfo  {journal} {Phys. Rev. D}\ }\textbf {\bibinfo {volume} {51}},\
  \bibinfo {pages} {2371} (\bibinfo {year} {1995})},\ \Eprint
  {http://arxiv.org/abs/hep-ph/9405371} {arXiv:hep-ph/9405371} \BibitemShut
  {NoStop}%
\bibitem [{\citenamefont {Meyer}(2010)}]{Meyer:2010}%
  \BibitemOpen
  \bibfield  {author} {\bibinfo {author} {\bibfnamefont {H.~B.}\ \bibnamefont
  {Meyer}},\ }\href@noop {} {\bibfield  {journal} {\bibinfo  {journal} {JHEP 04
  (2010) 099}\ } (\bibinfo {year} {2010})},\ \Eprint
  {http://arxiv.org/abs/arXiv:1002.3343} {arXiv:arXiv:1002.3343} \BibitemShut
  {NoStop}%
\bibitem [{\citenamefont {Aarts}\ \emph {et~al.}(2017)\citenamefont {Aarts},
  \citenamefont {Allton}, \citenamefont {De~Boni}, \citenamefont {Hands},
  \citenamefont {J\"ager}, \citenamefont {Praki},\ and\ \citenamefont
  {Skullerud}}]{Aarts:2017rrl}%
  \BibitemOpen
  \bibfield  {author} {\bibinfo {author} {\bibfnamefont {G.}~\bibnamefont
  {Aarts}}, \bibinfo {author} {\bibfnamefont {C.}~\bibnamefont {Allton}},
  \bibinfo {author} {\bibfnamefont {D.}~\bibnamefont {De~Boni}}, \bibinfo
  {author} {\bibfnamefont {S.}~\bibnamefont {Hands}}, \bibinfo {author}
  {\bibfnamefont {B.}~\bibnamefont {J\"ager}}, \bibinfo {author} {\bibfnamefont
  {C.}~\bibnamefont {Praki}}, \ and\ \bibinfo {author} {\bibfnamefont {J.-I.}\
  \bibnamefont {Skullerud}},\ }\href {\doibase 10.1007/JHEP06(2017)034}
  {\bibfield  {journal} {\bibinfo  {journal} {JHEP}\ }\textbf {\bibinfo
  {volume} {06}},\ \bibinfo {pages} {034} (\bibinfo {year} {2017})},\ \Eprint
  {http://arxiv.org/abs/1703.09246} {arXiv:1703.09246 [hep-lat]} \BibitemShut
  {NoStop}%
\bibitem [{\citenamefont {et~al.}(2019)}]{Galassi:2019}%
  \BibitemOpen
  \bibfield  {author} {\bibinfo {author} {\bibfnamefont {M.~G.}\ \bibnamefont
  {et~al.}},\ }\href@noop {} {\  (\bibinfo {year} {2019})}\BibitemShut
  {NoStop}%
\bibitem [{\citenamefont {Fousse}\ \emph {et~al.}(2007)\citenamefont {Fousse},
  \citenamefont {Hanrot}, \citenamefont {Lef\`{e}vre}, \citenamefont
  {P\'{e}lissier},\ and\ \citenamefont {Zimmermann}}]{10.1145/1236463.1236468}%
  \BibitemOpen
  \bibfield  {author} {\bibinfo {author} {\bibfnamefont {L.}~\bibnamefont
  {Fousse}}, \bibinfo {author} {\bibfnamefont {G.}~\bibnamefont {Hanrot}},
  \bibinfo {author} {\bibfnamefont {V.}~\bibnamefont {Lef\`{e}vre}}, \bibinfo
  {author} {\bibfnamefont {P.}~\bibnamefont {P\'{e}lissier}}, \ and\ \bibinfo
  {author} {\bibfnamefont {P.}~\bibnamefont {Zimmermann}},\ }\href {\doibase
  10.1145/1236463.1236468} {\bibfield  {journal} {\bibinfo  {journal} {ACM
  Trans. Math. Softw.}\ }\textbf {\bibinfo {volume} {33}},\ \bibinfo {pages}
  {13–es} (\bibinfo {year} {2007})}\BibitemShut {NoStop}%
\bibitem [{\citenamefont {Granlund}\ and\ \citenamefont {the GMP~development
  team}(2020)}]{Granlund:2020}%
  \BibitemOpen
  \bibfield  {author} {\bibinfo {author} {\bibfnamefont {T.}~\bibnamefont
  {Granlund}}\ and\ \bibinfo {author} {\bibnamefont {the GMP~development
  team}},\ }\href@noop {} {\  (\bibinfo {year} {2020})}\BibitemShut {NoStop}%
\bibitem [{\citenamefont {Korcyl}\ and\ \citenamefont
  {Bali}(2017)}]{Korcyl:2016ugy}%
  \BibitemOpen
  \bibfield  {author} {\bibinfo {author} {\bibfnamefont {P.}~\bibnamefont
  {Korcyl}}\ and\ \bibinfo {author} {\bibfnamefont {G.~S.}\ \bibnamefont
  {Bali}},\ }\href {\doibase 10.1103/PhysRevD.95.014505} {\bibfield  {journal}
  {\bibinfo  {journal} {Phys. Rev. D}\ }\textbf {\bibinfo {volume} {95}},\
  \bibinfo {pages} {014505} (\bibinfo {year} {2017})},\ \Eprint
  {http://arxiv.org/abs/1607.07090} {arXiv:1607.07090 [hep-lat]} \BibitemShut
  {NoStop}%
\bibitem [{\citenamefont {Dalla~Brida}\ \emph {et~al.}(2019)\citenamefont
  {Dalla~Brida}, \citenamefont {Korzec}, \citenamefont {Sint},\ and\
  \citenamefont {Vilaseca}}]{DallaBrida:2018tpn}%
  \BibitemOpen
  \bibfield  {author} {\bibinfo {author} {\bibfnamefont {M.}~\bibnamefont
  {Dalla~Brida}}, \bibinfo {author} {\bibfnamefont {T.}~\bibnamefont {Korzec}},
  \bibinfo {author} {\bibfnamefont {S.}~\bibnamefont {Sint}}, \ and\ \bibinfo
  {author} {\bibfnamefont {P.}~\bibnamefont {Vilaseca}},\ }\href {\doibase
  10.1140/epjc/s10052-018-6514-5} {\bibfield  {journal} {\bibinfo  {journal}
  {Eur. Phys. J. C}\ }\textbf {\bibinfo {volume} {79}},\ \bibinfo {pages} {23}
  (\bibinfo {year} {2019})},\ \Eprint {http://arxiv.org/abs/1808.09236}
  {arXiv:1808.09236 [hep-lat]} \BibitemShut {NoStop}%
\bibitem [{\citenamefont {Quenouille}(1956)}]{Quenouille:1956:NBE}%
  \BibitemOpen
  \bibfield  {author} {\bibinfo {author} {\bibfnamefont {M.~H.}\ \bibnamefont
  {Quenouille}},\ }\href@noop {} {\ \textbf {\bibinfo {volume} {43}},\ \bibinfo
  {pages} {353} (\bibinfo {year} {1956})}\BibitemShut {NoStop}%
\bibitem [{\citenamefont {Gattringer}\ and\ \citenamefont
  {Lang}(2010)}]{Gattringer:2010zz}%
  \BibitemOpen
  \bibfield  {author} {\bibinfo {author} {\bibfnamefont {C.}~\bibnamefont
  {Gattringer}}\ and\ \bibinfo {author} {\bibfnamefont {C.~B.}\ \bibnamefont
  {Lang}},\ }\href {\doibase 10.1007/978-3-642-01850-3} {\emph {\bibinfo
  {title} {{Quantum chromodynamics on the lattice}}}},\ Vol.\ \bibinfo {volume}
  {788}\ (\bibinfo  {publisher} {Springer},\ \bibinfo {address} {Berlin},\
  \bibinfo {year} {2010})\BibitemShut {NoStop}%
\bibitem [{\citenamefont {Bouma}\ \emph {et~al.}(2022)\citenamefont {Bouma},
  \citenamefont {Bali}, \citenamefont {Collins},\ and\ \citenamefont
  {S\"oldner}}]{Bouma:2022teu}%
  \BibitemOpen
  \bibfield  {author} {\bibinfo {author} {\bibfnamefont {S.}~\bibnamefont
  {Bouma}}, \bibinfo {author} {\bibfnamefont {G.}~\bibnamefont {Bali}},
  \bibinfo {author} {\bibfnamefont {S.}~\bibnamefont {Collins}}, \ and\
  \bibinfo {author} {\bibfnamefont {W.}~\bibnamefont {S\"oldner}} (\bibinfo
  {collaboration} {RQCD}),\ }\href {\doibase 10.22323/1.396.0548} {\bibfield
  {journal} {\bibinfo  {journal} {PoS}\ }\textbf {\bibinfo {volume}
  {LATTICE2021}},\ \bibinfo {pages} {548} (\bibinfo {year} {2022})},\ \Eprint
  {http://arxiv.org/abs/2206.04178} {arXiv:2206.04178 [hep-lat]} \BibitemShut
  {NoStop}%
\end{thebibliography}%

\end{document}